\UseRawInputEncoding
\documentclass[10pt,twocolumn,prc,showkeys]{revtex4}
\usepackage{slashed,graphicx}
\tighten

\begin{document}

\title{Translating neutron star observations to nuclear symmetry energy via \\
artificial neural networks}

\author{Plamen~G.~Krastev}
\affiliation{Harvard University, Faculty of Arts and Sciences, Research Computing, 52 Oxford Street, Cambridge, MA 02138, U.S.A.}
\email[Plamen G. Krastev: ]{plamenkrastev@fas.harvard.edu}


\date{\today}

\begin{abstract}
One of the most significant challenges involved in efforts to understand the equation of state of dense 
neutron-rich matter is the uncertain density dependence of the nuclear symmetry energy. In particular, 
the nuclear symmetry energy is still rather poorly constrained, especially at high densities. On the other hand, 
detailed knowledge of the equation of state is critical for our understanding of many important phenomena in the 
nuclear terrestrial laboratories and the cosmos. Because of its broad impact, pinning down the density dependence 
of the nuclear symmetry energy has been a long-standing goal of both nuclear physics and astrophysics. Recent 
observations of neutron stars, in both electromagnetic and gravitational-wave spectra, have already constrained 
significantly the nuclear symmetry energy at high densities. The next generation of telescopes and gravitational-wave 
observatories will provide an unprecedented wealth of detailed observations of neutron stars, which will improve 
further our knowledge of the density dependence of nuclear symmetry energy, and the underlying equation of state of
dense neutron-rich matter. Training deep neural networks to learn a computationally efficient representation 
of the mapping between astrophysical observables of neutron stars, such as masses, radii, and tidal deformabilities, 
and the nuclear symmetry energy allows its density dependence to be determined reliably and accurately. In this work,
we use a deep learning approach to determine the nuclear symmetry energy as a function of density directly from
observational neutron star data. We show, for the first time, that artificial neural networks can precisely reconstruct
the nuclear symmetry energy from a set of available neutron star observables, such as masses and radii as
measured by, e.g., the NICER mission, or masses and tidal deformabilities as measured by the LIGO/VIRGO/KAGRA
gravitational-wave detectors. These results demonstrate the potential of artificial neural networks to reconstruct
the symmetry energy and the equation of state directly from neutron star observational data, and emphasize
the importance of the deep learning approach in the era of multi-messenger astrophysics.
\end{abstract}

\maketitle

\section{Introduction}\label{sec1}

Understanding the equation of state (EOS) of dense neutron-rich matter in terms of the fundamental
interactions between its constituents is an extraordinarily challenging problem and represents a key outstanding 
question in modern physics and astrophysics~\cite{NAP2011,NAP2012,USLongRangePlan2015}. Due to its broad 
ramifications for many important phenomena, ranging from understanding the heavy ion collision dynamics in nuclear 
laboratories to the most violent cosmic events, such as binary neutron star (BNS) mergers and supernovae, to gravitational 
waves, the determination of the EOS of dense matter has been a major shared goal of both the nuclear physics (see, e.g., Refs.
\cite{EPJA2014,LiUniverse2021,Science2002,PhysRep2005v1,PhysRep2005v2,PRC2012,PPNP2016,NPN2017,PPNP2018,BurgioVidanaUniverse2020}) 
and astrophysics (see, e.g., Refs. \cite{Lattimer2001,Lattimer2016,Watts2016,Ozel2016,Oertel2017,Baiotti2019,EPJA2019,Weber2007,Alford2019,Capano2020,Blaschke2020,Chatziioannou2020,Annala2018,Kievsky2018,Landry2020,Dietrich2020,Stone2021,Li2020,Burgio:2021vgk,Burgio:2021bzy,KrastevJPG2019,Raithel:2019ejc})
communities. It has been a primary scientific thrust for establishing key research facilities in astrophysics \cite{NAP2011} and
nuclear physics \cite{NAP2012}, such as large ground-based telescopes, advanced X-ray space-borne observatories, the Neutron 
Star Interior Composition Explorer (NICER) \cite{NICER2017}, LIGO/VIRGO/KAGRA \cite{aLIGO2015,VIRGO:2014yos,KAGRA2019} 
gravitational-wave detectors, and all advanced radioactive beam laboratories around the globe.       

In cold neutron star matter, the nucleonic component of the EOS can be written in terms of the energy
per nucleon $\rho$ as \cite{Bombaci1991}
\begin{equation}\label{Eq.1}
E(\rho,\delta) = E_{SNM}(\rho)+E_{\rm{sym}}(\rho)\delta^2,
\end{equation}
where $E_{SNM}(\rho)=E(\rho,0)$ is the energy per nucleon of symmetric nuclear matter (SNM), $E_{\rm{sym}}(\rho)$ is the
symmetry energy, and $\delta=(\rho_{\rm{n}}-\rho_{\rm{p}})/\rho$ is the isospin asymmetry, with $\rho_{\rm{n}}$, $\rho_{\rm{p}}$,
and $\rho = \rho_{\rm{n}} + \rho_{\rm{p}}$ being the neutron, proton, and total density, respectively. Presently, the EOS of cold
nuclear matter under extreme conditions of density, pressure, and/or isospin asymmetry still remains rather uncertain and
theoretically controversial, particularly at supra-saturation densities, mainly due to the poorly known high-density behavior of
the nuclear symmetry energy $E_{sym}(\rho)$ \cite{EPJA2014,LiUniverse2021}.

To determine the EOS from first principles, we need to solve quantum chromodynamics (QCD), which is the fundamental theory
of strong interactions. However, at present, model-independent results are only available for a rather limited density
range. At low densities of $\rho \sim [1 - 2]\rho_0$~\endnote{$\rho_0$ = 0.16 fm$^{-1}$ is the saturation density of symmetric nuclear matter.}, 
we can use \textit{ab initio} approaches together with nuclear interaction
derived from Chiral Effective Theory ($\chi$EFT) with controlled uncertainty estimates~\cite{Hebeler2010,Tews2013,Holt2013,Hagen2014,Roggero2014,Machleidt2011,Wlazlowski2014,Tews2018,Drischler2020,Drischler2021}.
At asymptotically high densities of $\rho  \gtrsim 50~\rho_0$, perturbative QCD calculations converge and provide reliable 
results \cite{Freedman1977v1,Freedman1977v2,Baluni1978,Kurkela2010,Fraga2014,Gorda2018,Ghiglieri2020}. At intermediate 
densities of $\rho \sim [2 - 10]\rho_0$, however, there are still no reliable QCD predictions \cite{Fujimoto:2021zas}. To derive the EOS 
from QCD in the intermediate-density region, one needs to develop non-perturbative approaches, such as the Monte Carlo simulation 
of QCD on a lattice (lattice QCD), but the application of these methods to systems at finite densities is hindered by the notorious 
sign problem; see, e.g., Ref. \cite{Aarts2016}. Therefore, at intermediate densities, the EOS construction still relies on 
phenomenological approaches employing a variety of many-body methods and effective interactions, such as the relativistic 
mean field theory, and density functionals based on the Skyrme, Gogny, or Similarity Renormalization Group (SRG) 
evolved interactions. 

Meanwhile, we have witnessed extraordinary progress in efforts to constrain the high-density EOS from both nuclear
laboratory experiments with radioactive beams, and multi-messenger astrophysical (MMA) observations of neutron stars. 
In particular, extensive analyses of experimental data of heavy-ion reactions from intermediate to relativistic energies, 
especially various forms of nucleon collective flow and the kaon production, have already constrained significantly the EOS 
of SNM up to approximately $4.5~\rho_0$; see, e.g., Ref.~\cite{Science2002}. In addition, thanks to the great efforts and collaboration 
of both the nuclear physics and astrophysics communities, significant progress has been made in the last two decades in 
constraining  the symmetry energy around, and below, nuclear matter saturation density using results from both  astrophysical 
observations and terrestrial nuclear experiments; see, e.g., Refs. 
\cite{LiUniverse2021,PRC2012,PPNP2016,NPN2017,LiPhysRep2008,LiPLB2013,Horowitz2014,LattimerEPJA2014}. 
However, the poorly known density dependence of the nuclear symmetry energy $E_{sym}(\rho)$ at supra-saturation densities, 
and the possible hadron-to-quark phase transition, still remain the most uncertain aspects of the EOS of dense matter~
\cite{EPJA2014,LiUniverse2021,Weber2007,Alford2019,Blaschke2020}. Furthermore, the appearance of various new particles, 
such as hyperons and resonances, is also strongly dependent upon the high-density trend of $E_{sym}(\rho)$~\cite{Drago2014,Cai2015,Zhu2016,Sahoo2018,LiJJ2018,LiJJ2019,Ribes2019,Raduta2020,Raduta2021,Thapa2021,Sen2021,Jiang2012,Providencia2019,Vidana2018,Choi2021,Fortin2021}. 
Because, above the hadron-to-quark transition density, the nuclear symmetry energy would naturally
lose its physical meaning, it is critical to determine simultaneously both the high-density behavior of the symmetry energy
and the detailed properties of the hadron-to-quark phase transition, analyzing combined data from astrophysical observations
and nuclear laboratory experiments \cite{LiUniverse2021}.

Recent MMA observations of neutron stars provide unique means to probe the high-density EOS and, in particular, the
$E_{sym}(\rho)$ at densities currently inaccessible in the nuclear laboratories. Moreover, these new advances in neutron star (NS) 
observations have opened an alternative pathway for the model-independent extraction of the symmetry energy, and the EOS, 
via statistical approaches (see, e.g., Refs. \cite{Raithel:2019ejc,OzelPRD2010,Steiner2010,Steiner2013,Raithel2016,Raithel2017,Essick2020}). 
These observations include the Shapiro delay measurements of massive $\sim$$2M_{\odot}$  pulsars \cite{Demorest2010,Antoniadis2013,Cromartie2020}, 
the radius measurement of quiescent low-mass X-ray binaries and thermonuclear bursters \cite{OzelPRD2010,Steiner2010,Steiner2013,OzelApJ2016,Bogdanov2016},
the X-ray timing measurements of pulsars by the NICER mission \cite{Riley2019,Miller2019}, and the detection and inference of 
gravitational waves from compact binary mergers involving NSs by the LIGO/VIRGO/KAGRA collaboration \cite{BNS2017,BNS2019,NSBH2021}. 
Typical NS observables include mass $M$, radius $R$, moment of inertia $I$, quadrupole moment $Q$, dimensionless tidal deformability 
$\Lambda$ (and derivatives, e.g., Love number $k_2$ and tidal deformability $\lambda$), and compactness $M/R$. Specifically, 
the NICER mission aims at the compactness $M/R$ of NSs by measuring the gravitational lensing of the thermal emission from the 
stellar surface. On the other hand, gravitational-wave (GW) observations of BNS and neutron star--black hole (NSBH) mergers provide 
information on the tidal disruption of the star in the presence of its companion, quantified by the tidal deformability parameter 
$\lambda$. Some of these NS observables are related via EOS-independent universal relations, such as the well-known I-Love-Q relation, which 
relates $I$, $k_2$, and $Q$ \cite{YagiSci2013,YagiPRD2013}.

There is a plethora of diverse statistical approaches to construct the most probable EOS from NS observational
data, with the Bayesian inference \cite{Raithel:2019ejc,OzelPRD2010,Steiner2010,Steiner2013,Raithel2016,Raithel2017} 
as the most commonly used technique at the present. There are also other methods, such as those based on the Gaussian 
processes, which are variants of the Bayesian inferences with nonparametric representation of the EOS; see, e.g., Ref. \cite{Essick2020}. 
Despite the significant effort to extract the genuine EOS from the NS astrophysical data, it is still unclear what the {\it true} dense 
matter EOS should look like, mainly due to the uncertainties of the assumed prior distributions in the Bayesian analyses \cite{Fujimoto:2021zas}. 
Therefore, the need arises for alternative approaches to construct the model-independent EOS. Recently, approaches based 
on deep neural networks (DNNs) have gained interest in the research community and have been extensively explored and applied 
in a wide range of scientific and technical domains. Deep learning (DL) algorithms, a subset of machine learning (ML), are highly 
scalable computational techniques with the ability to learn directly from raw data, employing artificial neurons arranged in stacked 
layers, named neural networks, and optimization methods based on gradient descent and back-propagation \cite{LeCun2015,Goodfellow2016}.
These techniques, especially with the aid of GPU computing, have proven to be highly successful in tasks such as image 
recognition \cite{He2016}, natural language processing \cite{Young2018}, and recently also emerged  as a new tool in 
engineering and scientific applications, alongside traditional High-Performance Computing (HPC) in the new 
field of Scientific Machine Learning \cite{Baker2019}. DL approaches have been already successfully applied in a wide range of
physics and astrophysics domains; see, e.g., Refs.~\cite{Pang2018,Mori2017,Porotti2019,Rem2019,Melko2019,Carleo2017,Shanahan:2018vcv,Liu2021,Gomez:2021hqd,Villar2020,Schwartz2021}. 
In particular, DL has been applied in GW data analysis for the detection 
\cite{Gabbard2018,GeorgePRD2018,GeorgePLB2018,Gebhard2019,Wang2020,Lin:2020aps,Morales2021,Xia2021}, 
parameter estimation \cite{Chua2020,Green2021}, and denoising \cite{Wei2020} of GW signals from compact binary mergers. 
In previous works \cite{Krastev2020,Krastev2021}, we also pioneered the use of DL methods, specifically Convolutional Neural Network 
(CNN) \cite{Lecun1998} algorithms, for the detection and inference of GW signals from BNS mergers embedded in both
Gaussian and realistic LIGO noise. Several studies have also explored the DL approach as a tool to extract the dense matter EOS from 
NS observations \cite{Fujimoto:2021zas,Ferreira:2019bny,Morawski:2020izm,Traversi:2020dho,Fujimoto2020}.

In this work, we explore a DL approach to extract the nuclear symmetry energy $E_{sym}(\rho)$, the most uncertain
part of the EOS, directly from NS astrophysical data. Specifically, we train DNNs to map pairs of NS mass and radius 
$M-R$, or stellar mass and tidal deformability $M-\Lambda$, to $E_{sym}(\rho)$. We show, for the first time, that DL can be used 
to infer the nuclear symmetry energy directly from NS observational data accurately and reliably. Most importantly, we show that DL 
algorithms can be used successfully to construct a model-independent $E_{sym}(\rho)$ and therefore determine precisely the density 
dependence of the nuclear symmetry energy at supra-saturation densities. These results are a step towards achieving the goal of 
determining the EOS of dense neutron-rich matter, and emphasize the potential and importance of this DL approach in 
the MMA era, as an ever-increasing volume of NS observational data becomes available with the advent of the next generation of large 
telescopes and GW observatories.

This paper is organized as follows. After the introductory remarks in this section, in Section~\ref{sec2}, we discuss the main features and parameterization
of the EOS applied in this work. In Section~\ref{sec3}, we briefly recall the formalism for solving the structure equations of static NSs and
calculating the tidal deformability. In Section~\ref{sec4}, we discuss the DL approach used to map the NS observables to the nuclear symmetry energy.
We present our results in Section~\ref{sec5}. At the end, we conclude in Section~\ref{sec6} with a short summary and outlook on future investigations.

\textit{Conventions}: We use units in which $G = c = 1$.

\section{Equation of State}\label{sec2}

The EOS is the major ingredient for solving the NS structure equations and calculating global stellar properties, such as mass 
$M$, radius $R$, and dimensionless tidal deformability $\Lambda$. The most commonly used theoretical approaches to
determine the nuclear EOS fall into two major categories---phenomenological and microscopic methods. Phenomenological
approaches are based on effective interactions constructed to describe the ground state of finite nuclei and therefore applications
to systems at high isospin asymmetries must be considered with care \cite{Stone2007}. Moreover, at large densities, no experimental data are
available to constrain such interactions and therefore predictions based on these methods could be very different from the realistic behavior. 
Among the most used phenomenological approaches are methods based on Skyrme interactions \cite{Vautherin1972,Quentin1978} and 
relativistic mean-field (RMF) models~\cite{Boguta1977}. On the other hand, microscopic approaches start with realistic two-body
and three-body nucleon forces that describe accurately free-space nucleon scattering data and the deuteron properties. Such
interactions are either based on meson-exchange theory~\mbox{\cite{Machleidt1987,Nagelis1978},} or recent $\chi$EFT 
\cite{Machleidt2011,Weinberg1990,Weinberg1991,Epelbaum2009}. 
The major challenge for the many-body methods is the treatment of the short-range repulsive core of the nucleon--nucleon interaction, 
and this represents the difference among the available techniques. Among the most well-known microscopic many-body methods are 
the Brueckner--Hartree--Fock (BHF) approach \cite{Day1967} and its relativistic counterpart, the Dirac--Brueckner--Hartree--Fock (DBHF) 
theory \cite{Brockmann1990,Muther2017}, the variational approach \cite{Akmal1998}, the Quantum Monte Carlo technique and its 
derivatives~\mbox{\cite{Wiringa2000,Gandolfi2009}}, the self-consistent Green's function technique \cite{Kadanoff1962}, the $\chi$EFT~\mbox{\cite{Drischler2020},} 
and the $V_{low\; k}$ approach~\cite{Bogner2010}.

\begin{figure*}[t!]
\centering
\includegraphics[scale=0.46]{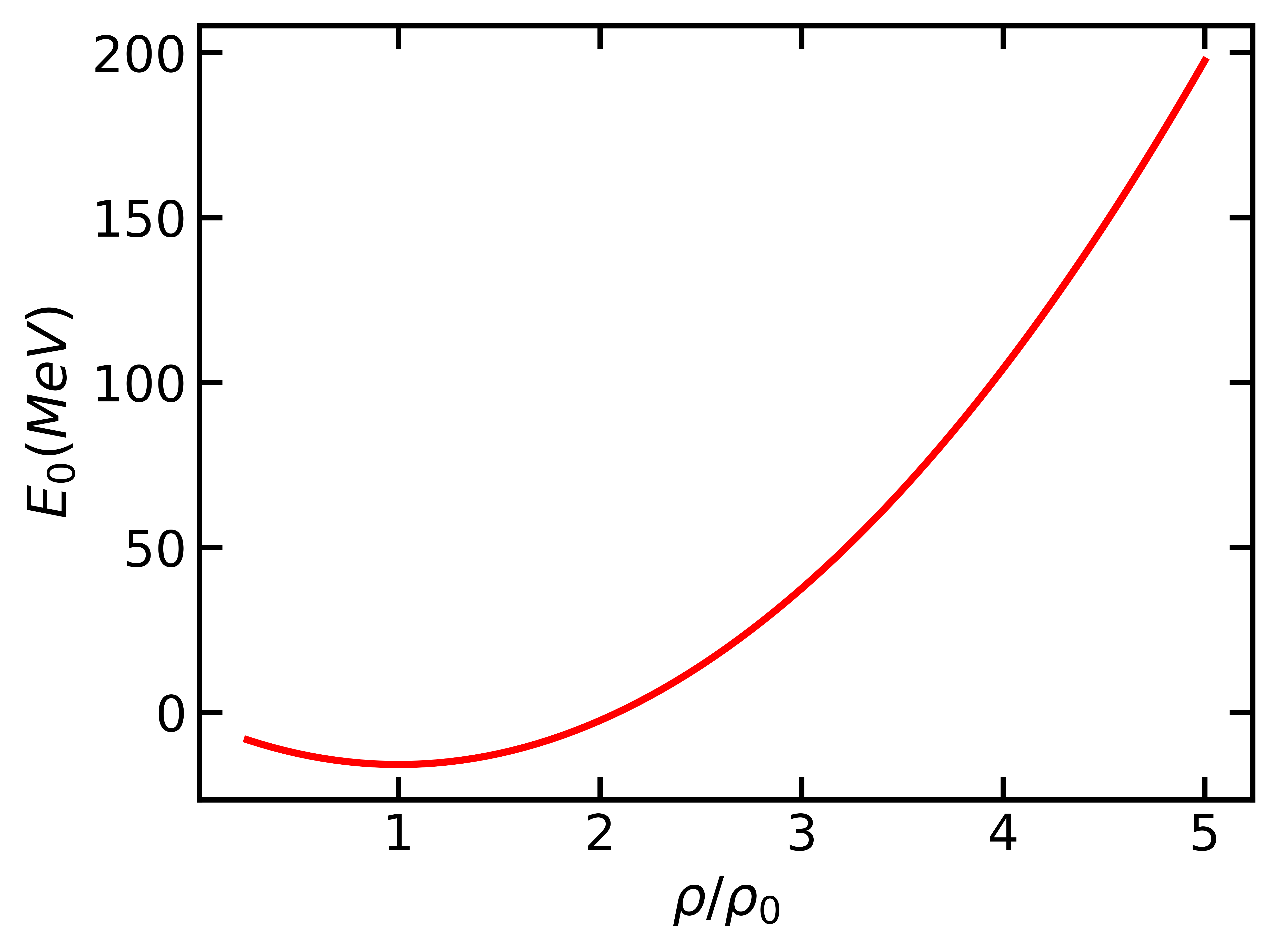}
\includegraphics[scale=0.46]{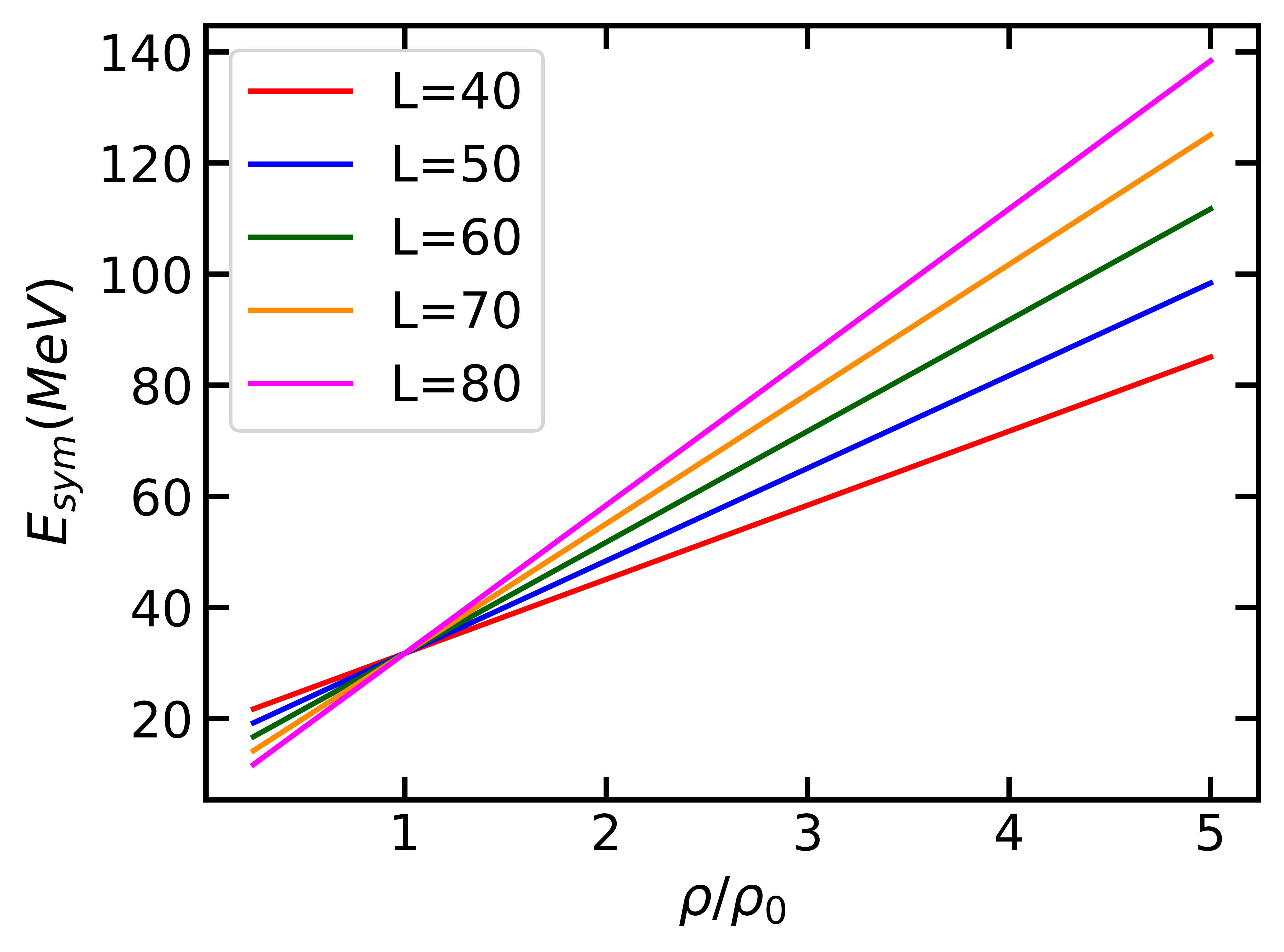}
\includegraphics[scale=0.46]{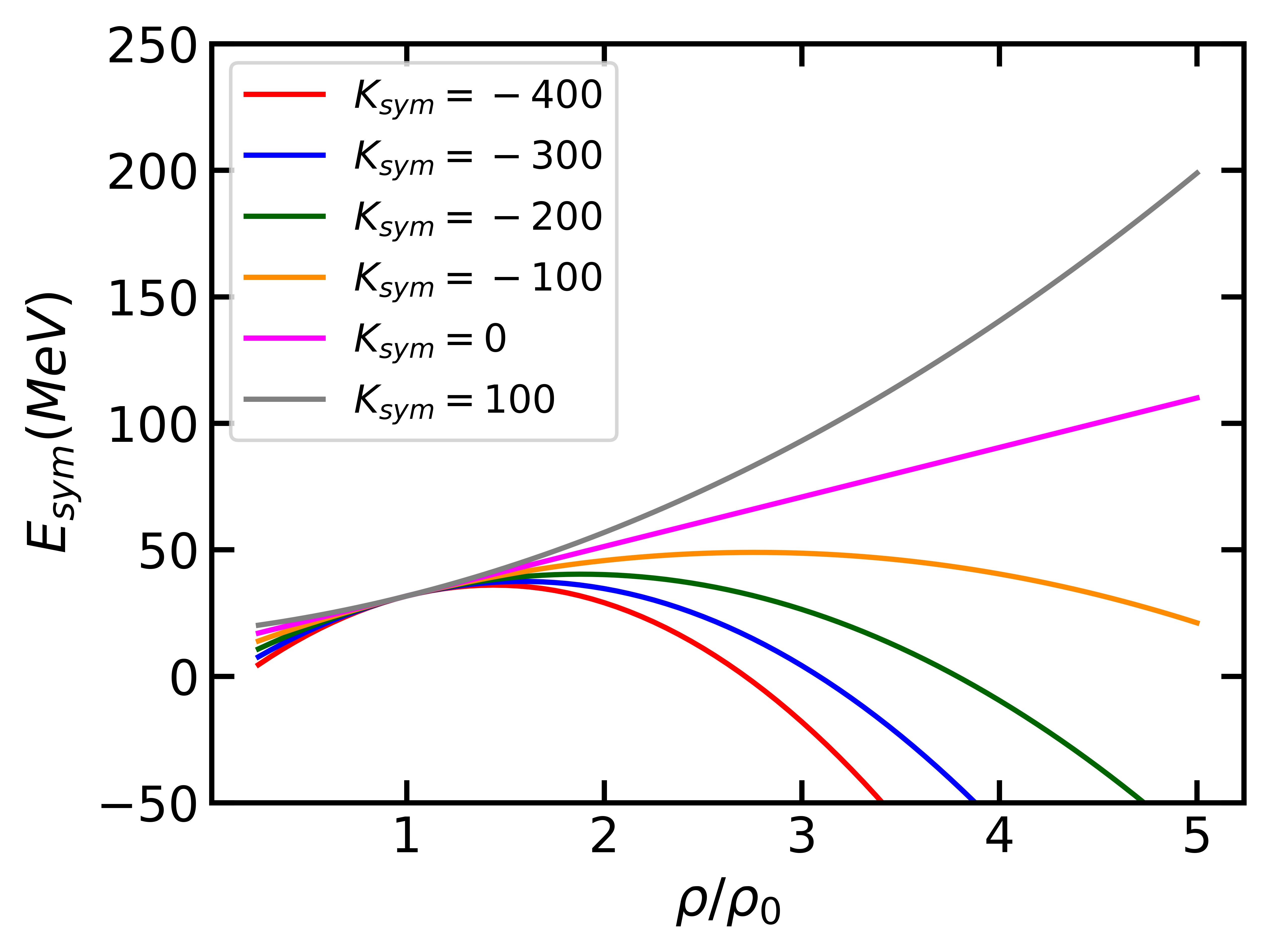}
\caption{{ (\textbf{Left window})} Energy per particle of SNM as a function of the reduced density $\rho/\rho_0$. The EOS of SNM is kept fixed
by setting all parameters in Equation~(\ref{Eq.1}). Specifically, we set $E_0=15.9$ MeV, $K_0=240$ MeV, and $J_0=0$ MeV. 
{ (\textbf{Middle window})}  Symmetry energy $E_{sym}$ as a function of $\rho/\rho_0$ for $L$ = 40, 50, 60, 70, 80 MeV, with $K_{sym}=0$ MeV.
{ (\textbf{Right window})} Same as the middle window but for $K_{sym}$ = $-$400, $-$300, $-$200, $-$100, 0, 100 MeV, with $L=58.7$ MeV. In both
middle and right windows, $S_0=31.7$ MeV and $J_{sym}=0$ MeV. See text for details.}\label{fig1}
\end{figure*}

Around saturation density $\rho_0$, the $E_{SNM}(\rho)$ and $E_{sym}(\rho)$ predicted by many-body theories can be Taylor 
expanded as
\begin{equation}\label{Eq.2}
E_{SNM}(\rho) = E_0 + \frac{K_0}{2}x^2 + \frac{J_0}{6}x^3,
\end{equation}
\begin{equation}\label{Eq.3}
E_{sym}(\rho) = S_0 + L x + \frac{K_{sym}}{2}x^2 + \frac{J_{sym}}{6}x^3,
\end{equation}
with $x\equiv(\rho-\rho_0)/3\rho_0$. The expansion coefficients in these expressions can be constrained by nuclear 
experiments and have the following meanings \cite{Vidana2009}: 
$E_0 \equiv E_{SNM}(\rho_0)$, 
$K_0\equiv[9\rho^2d^2E_{SNM}/d\rho^2]_{\rho_0}$, and 
$J_0\equiv[27\rho^3d^3E_{SNM}/d\rho^3]_{\rho_0}$ are the binding energy, incompressibility, and skewness of SNM; 
$S_0 \equiv E_{sym}(\rho_0)$, 
$L\equiv[3\rho dE_{sym}/d\rho]_{\rho_0}$, 
$K_{sym}\equiv[9\rho^2d^2E_{sym}/d\rho^2]_{\rho_0}$, and 
$J_{sym}\equiv[27\rho^3d^3E_{sym}/d\rho^3]_{\rho_0}$ are the magnitude, slope, curvature, and skewness of the symmetry 
energy at $\rho_0$. Currently, the most probable values of these parameters are as follows: 
$E_0 = -15.9\pm 0.4$ MeV, 
$K_0 = 240\pm 20$ MeV, 
$-300\le J_0 \le 400$ MeV, 
$S_0 = 31.7\pm 3.2$~MeV, 
$L = 58.7\pm 28.1$ MeV, 
$-400\le K_{sym}\le 100$ MeV, and 
$-200\le J_{sym}\le 800$~MeV; 
see e.g., Ref. \cite{Zhang2018}.
Although, at higher densities, the Taylor expansions diverge themselves~\cite{Cai2021}, Equations~(\ref{Eq.2}) and (\ref{Eq.3}) can also 
be viewed as parameterizations where, in principle, the parameters are left free \cite{Zhang2018}. In this respect, the above relations 
have dual meanings. Namely, for systems with low isospin asymmetries, they are Taylor expansions near the saturation density, 
while, for very neutron-rich systems at supra-saturation densities, they should be regarded as parameterizations \cite{Zhang2018}.
{For further discussion on the relationship between the \textit{Taylor expansions} and the parameterizations adopted in our
analysis, the reader is referred to, e.g., Ref. \cite{Zhang2018}.} 
These parameterizations are often used in \textit{metamodeling} of the NS EOS and have been applied previously, for instance, 
in solving the NS inverse-structure problem and constraining the high-density symmetry energy by astrophysical observations 
of NSs~\cite{Zhang2018,Zhang2019}. The NS EOS metamodel has been also applied in Bayesian  analyses to extract the most probable 
values of the high-density parameters of the EOS, where the posterior Probability Distribution Functions (PDFs) of the EOS 
parameters and their correlations are inferred directly from NS observational data \cite{Xie2019}. The parameterizations described
here have the advantage, over the widely used piecewise polytropes for directly parameterizing the pressure as a function
of energy or baryon density of NS matter, of keeping the isospin dependence of the EOS, and they explicitly retain information on 
the composition for the whole density range, without losing the ability to model a wide range of EOSs as predicted by
various many-body approaches. This feature of the metamodeling approach is particularly important for inferring the high-density 
symmetry energy parameters, or directly $E_{sym}(\rho)$, as it clearly separates the contribution of $E_{sym}(\rho)$ to the EOS.   

\begin{figure*}[t!]
\centering
\includegraphics[scale=0.5]{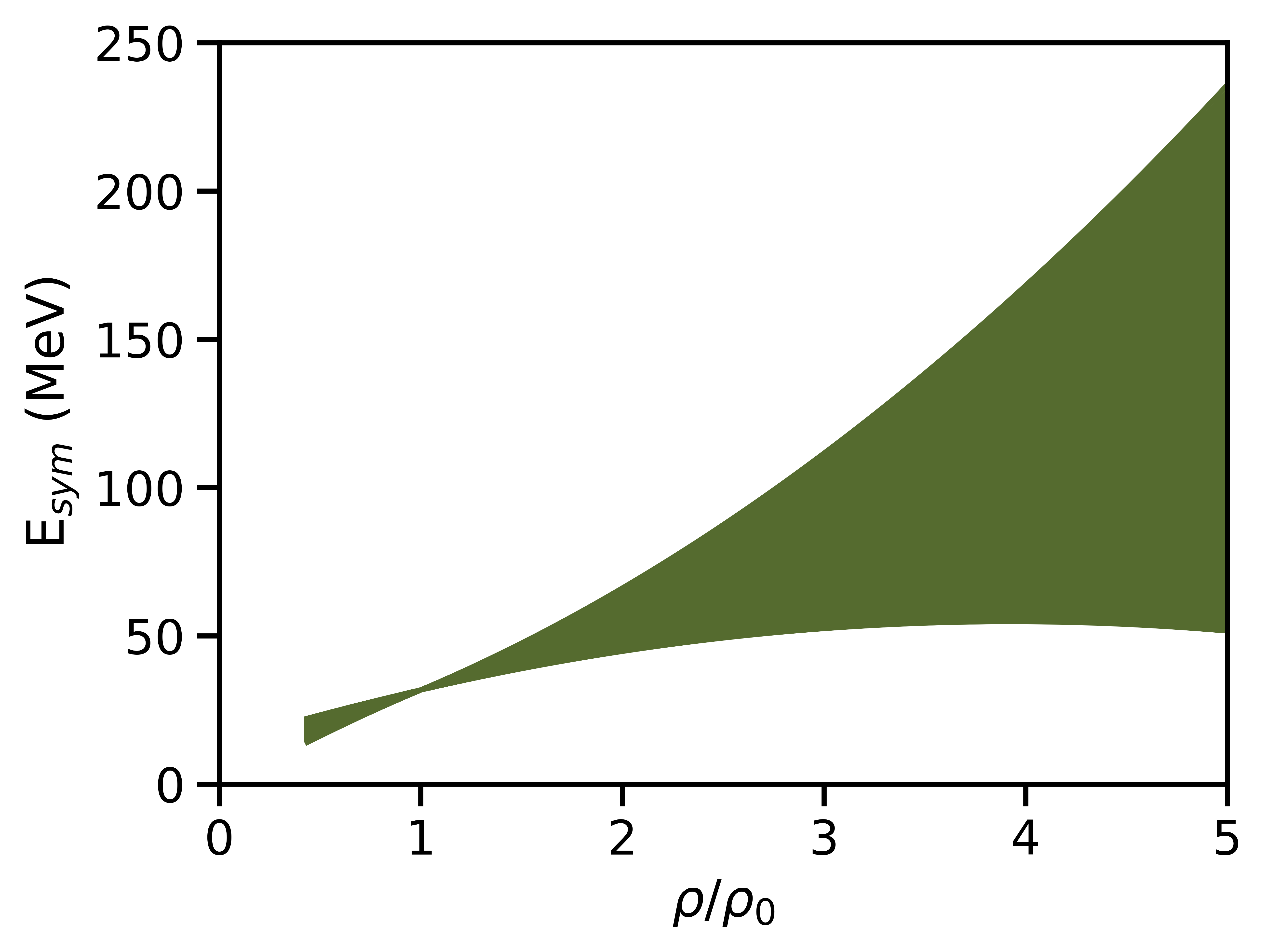}
\includegraphics[scale=0.5]{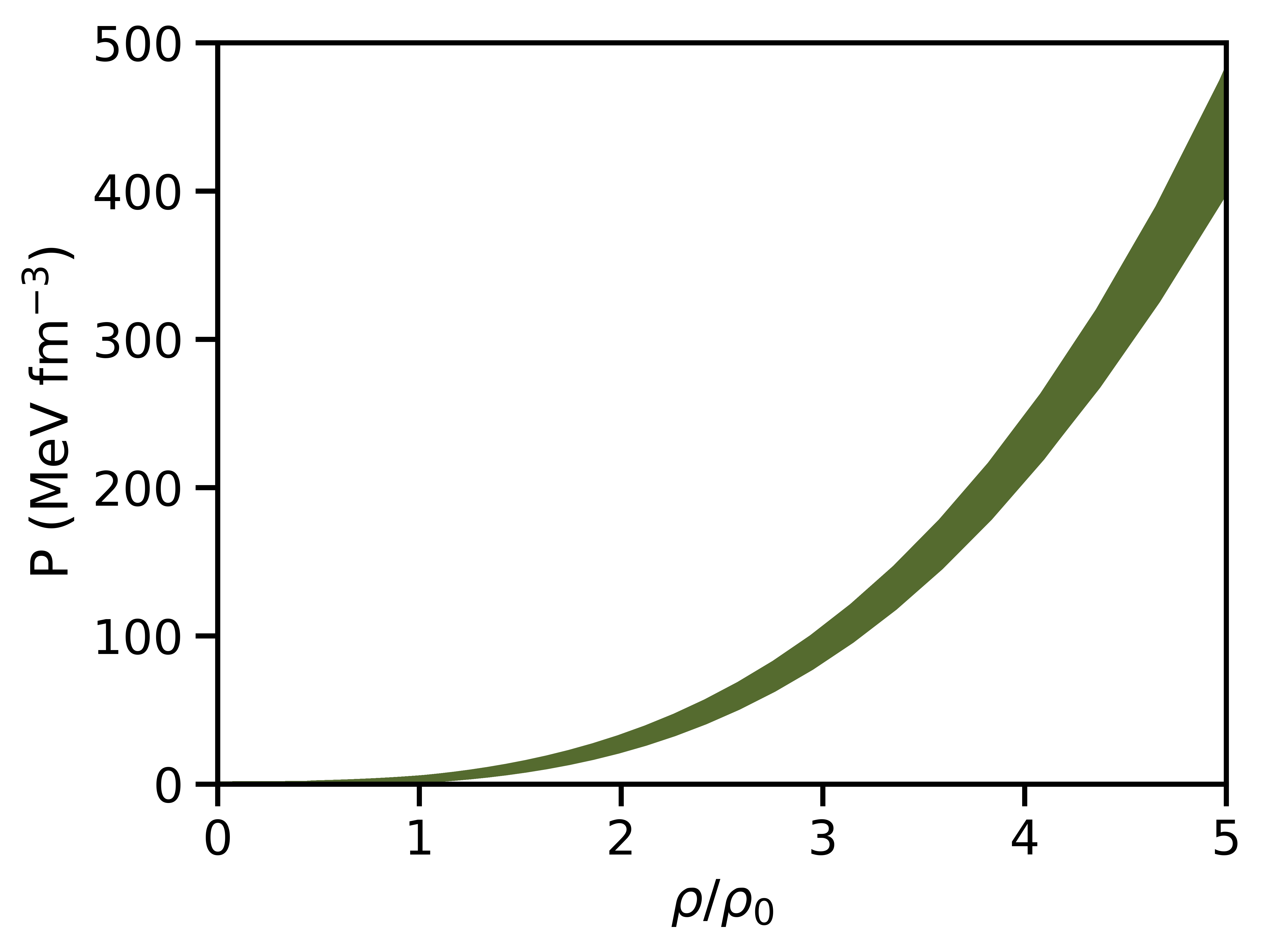}
\caption{Range of the nuclear symmetry energy $E_{sym}$ (\textbf{left window}) and pressure $P$ (\textbf{right window}). The $E_{sym}$
and $P$ are plotted as functions of the reduced density $\rho/\rho_0$.}\label{fig2}
\end{figure*}

For the purpose of our analysis, in the present study, we adopt the EOS metamodel described briefly above and, by varying the 
EOS parameters, generate a large number of EOSs and corresponding $M-R$, or $M-\Lambda$ sequences, by solving the 
NS structure equations. We assume a simple model of matter in the NS core consisting of protons, neutrons, electrons, and 
muons (the $npe\mu$-model) in $\beta$-equilibrium. With $E_{SNM}(\rho)$ and $E_{sym}(\rho)$, parameterized by 
Equations~(\ref{Eq.2}) and (\ref{Eq.3}), $E(\rho,\delta)$ is calculated through Equation~(\ref{Eq.1}). Then, the pressure of NS matter in 
$\beta$-equilibrium 
\begin{equation}\label{Eq.4}
P(\rho,\delta)=\rho^2\frac{d\epsilon(\rho,\delta)/\rho}{d\rho}
\end{equation}
can be computed from the energy density $\varepsilon(\rho,\delta)=\rho[E_n(\rho,\delta)+M_N]+\varepsilon_l(\rho,\delta)$, 
where $M_N$ is the average nucleon mass and $\varepsilon_l(\rho,\delta)$ is the lepton energy density. Details for calculating  
$\varepsilon_l(\rho,\delta)$ can be found in, e.g., Ref. \cite{Krastev2006}. Below approximately 0.07 $fm^{-3}$, the core EOS is 
supplemented by a crustal EOS, which is more suitable at lower densities. For the inner crust, we apply the EOS by 
Pethick et al. \cite{Pethick1995} and, for the outer crust, the one by Haensel and Pichon \cite{Haensel1994}. 

We use Equations (\ref{Eq.2}) and (\ref{Eq.3}) as parameterizations, together with the parabolic approximation of the nucleonic EOS 
Equation~(\ref{Eq.1}), and fix $E_0$, $K_0$, and $S_0$ at their most probable currently known values from nuclear
laboratory experiments and/or nuclear theories. Since the main focus of this analysis is specifically on extracting $E_{sym}(\rho)$ 
from NS observables, we fix the SNM EOS, $E_{SNM}(\rho)$, by setting $J_0=0$ MeV, and vary only $E_{sym}(\rho)$. The $J_0$ 
parameter controls the stiffness of the SNM EOS and in turn the maximum NS mass $M_{max}$ of the resultant stellar models. 
The maximum NS mass of 2.14~M$_{\odot}$ observed so far~\cite{Cromartie2020} requires $J_0$ to be larger than $-$200 MeV, 
depending slightly on the symmetry energy parameters~\cite{Li:2020ass}. At present, the predicted range of $J_0$ still has
relatively large uncertainties~\cite{Li:2020ass}, which also {partially} justifies our choice of setting $J_0=0$. 
With $E_{SNM}(\rho)$ kept fixed, the EOS is therefore solely determined by $E_{sym}(\rho)$. The EOS of SNM $E_{SNM}(\rho)$ is shown in the left window of 
Figure~\ref{fig1}. For the purpose of our analysis, we also set the symmetry energy skewness parameter $J_{sym}=0$ MeV. This 
choice is also partially justified by the very large uncertainty range of $J_{sym}$ at present \cite{Li:2020ass}, {but the main reason for
setting $J_{sym}=0$ is to simplify our problem. In following works, we plan to consider the effect of both $J_0$ and $J_{sym}$. This would
allow for the modeling of a wider class of EOSs as predicted by various many-body approaches, and models of the nuclear interaction. Together
with using realistic NS astrophysical observations, this data-driven approach would allow for extracting realistic symmetry energies, and in turn the EOS.} 
With {the above} choices, we  {subsequently} vary the symmetry energy parameters $L$ and $K_{sym}$ 
to generate many samples of $E_{sym}(\rho)$, and the EOS. The effect of varying the $L$ and $K_{sym}$ parameters on the symmetry energy is 
illustrated in the middle and right windows of Figure~\ref{fig1}. While, in principle, these parameters are absolutely free, the asymptotic boundary conditions 
of the EOS near $\rho_0$ and $\delta=0$ provide some prior knowledge of the ranges of these parameters. The ranges of $L$ and $K_{sym}$ are 
further restricted by imposing the requirement that the EOSs satisfy causality, and the resultant NS models can support a maximal mass of at least 
2.14~M$_{\odot}$. The ranges of the symmetry energy $E_{sym}(\rho)$ and pressure $P$ satisfying all constraints are shown in Figure~\ref{fig2}.

\section{Neutron Star Structure Equations and Tidal Deformability}\label{sec3}

In this section, we briefly review the formalism for calculating the NS mass $M$, radius $R$, and tidal deformability $\lambda$.
For a spherically symmetric relativistic star, the Einstein's field equations reduce to the familiar 
Tolaman--Oppenheimer--Volkoff (TOV) \cite{Oppenheimer1939} equation:
\begin{eqnarray} \label{Eq.5} 
\frac{dp(r)}{dr} && = -\frac{\varepsilon(r)m(r)}{r^2}\left[1+\frac{p(r)}{\varepsilon(r)}\right] \nonumber \\
                        && \times\left[1+\frac{4\pi{r^3}p(r)}{m(r)}\right]\left[1-\frac{2m(r)}{r}\right]^{-1}, 
\end{eqnarray}
where the gravitational mass within a sphere of radius $r$ is determined by
\begin{equation}
\frac{dm(r)}{dr}=4\pi\varepsilon(r)r^{2}.    \label{Eq.6}
\end{equation}

To proceed with the solution of the above equations, one needs to provide the EOS of stellar matter in the form
$p(\varepsilon)$. Starting from some central energy density $\varepsilon_c=\varepsilon(r=0)$ at the center
of the star, with the initial condition $m(0)=0$, Equations~(\ref{Eq.5}) and (\ref{Eq.6}) can be integrated until $p$ vanishes,
signifying that the edge of the star has been reached. Some care should be taken at $r=0$ since, as seen above, the 
TOV equation is singular there. The point $r=R$ where $p_0$ vanishes defines the NS radius and 
$M=m(R)=4\pi\int^R_0\varepsilon(r')r'^2dr'$ determines the NS gravitational mass. 

\begin{figure*}[t!]
\centering
\includegraphics[scale=0.5]{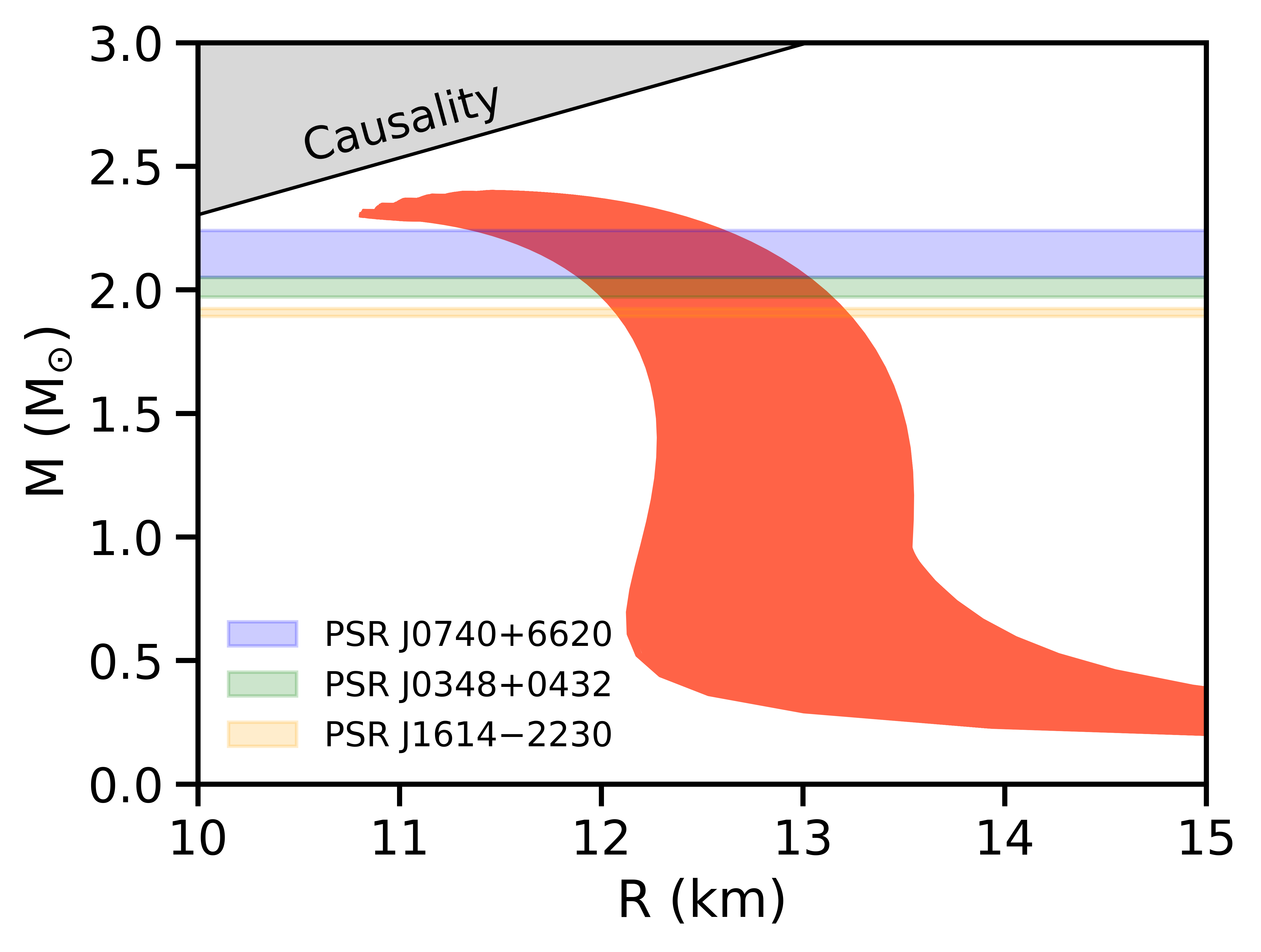}
\includegraphics[scale=0.5]{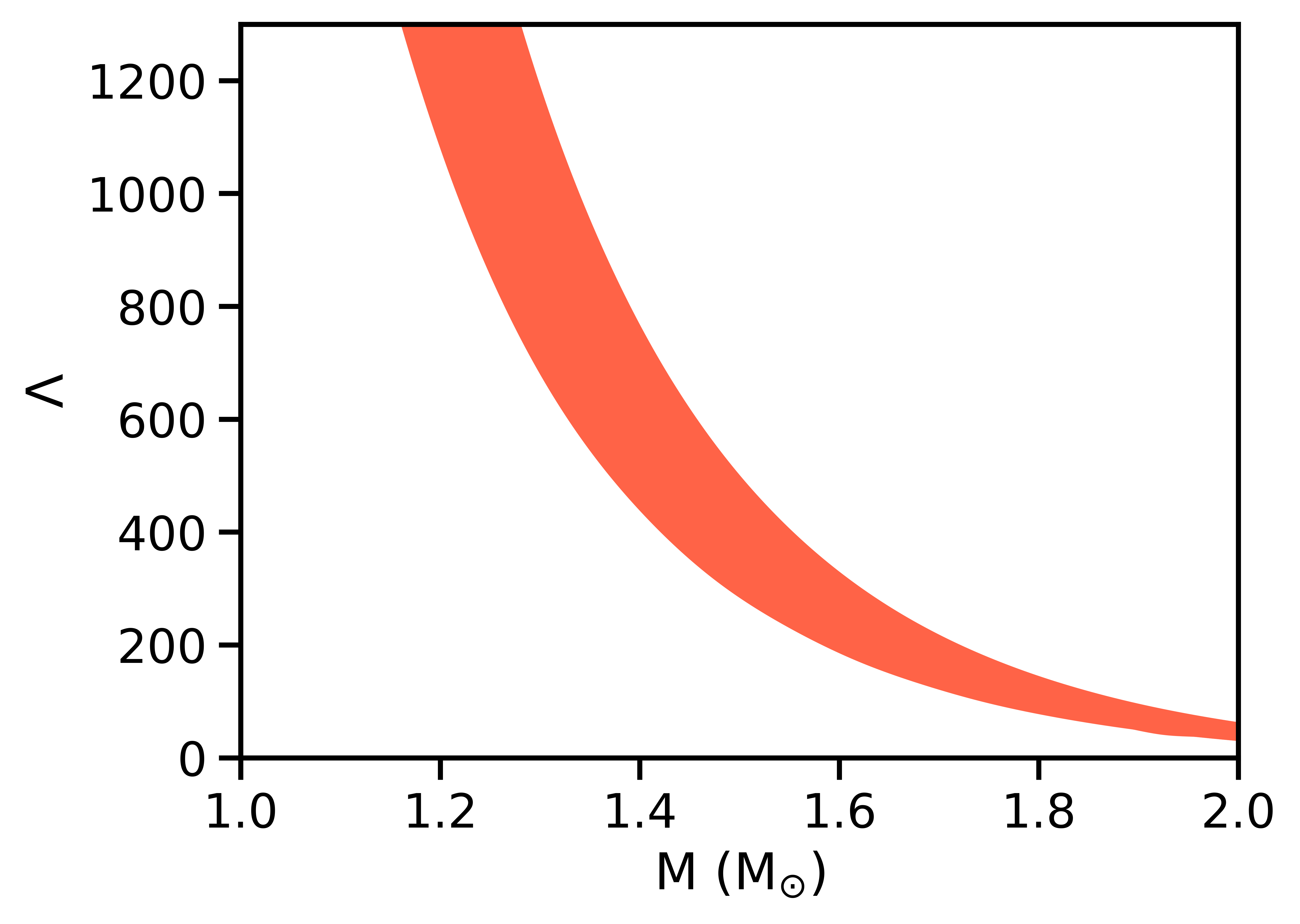}
\caption{Range of mass--radius relation (left window) and dimensionless tidal deformability $\Lambda$ (right window) for the EOSs
considered in this study. $\Lambda$ is plotted as a function of stellar mass $M$. The mass ranges of the three heaviest pulsars known
at present \cite{Demorest2010,Antoniadis2013,Cromartie2020} are indicated in the left window.}\label{fig3}
\end{figure*}

For a given EOS, there is a unique relationship between the stellar mass and the central density $\varepsilon_c$. 
Thus, for a particular EOS, there is a unique sequence of NSs parameterized by the central density (or equivalently 
the central pressure $p_c=p(0)$). The range of the $M-R$ relation computed with the EOSs considered in this work
is shown in the left window of Figure~\ref{fig3}.

The tidal deformability $\lambda$ is a parameter quantifying the tidal deformation effects experienced by
NSs in coalescing binary systems during the early stages of an inspiral. This parameter is defined 
as~\cite{Hinderer:2009ca,Flanagan:2007ix,Damour:2009vw}
\begin{equation}\label{Eq.7}
\lambda = -\frac{Q_{ij}}{\mathcal{E}_{ij}},
\end{equation}
where $Q_{ij}$ is the induced mass quadruple moment of an NS in the gravitational tidal field $\mathcal{E}_{ij}$
of its companion. The tidal deformability can be expressed in terms of the NS radius, $R$, and
dimensionless tidal Love number, $k_2$, as
\begin{equation}\label{Eq.8}
\lambda = \frac{2}{3}k_2 R^5.
\end{equation}

The tidal Love number $k_2$ is calculated using the following
expression~\cite{Hinderer:2007mb,Postnikov:2010yn}:
\begin{eqnarray}\label{Eq.9}
k_2(\beta, y_R) & = & \frac{8}{5}\beta^5(1-2\beta)^2 [2-y_R+2\beta(y_R-1)]                \nonumber \\
                          &\times& \{2\beta[6-3y_R+3\beta(5y_R-8)]                                         \nonumber \\
                          &+& 4\beta^3[13-11y_R + \beta(3y_R-2)                                             \nonumber \\
                          &+&2\beta^2(1+y_R)] +3(1-2\beta)^2 [2-y_R                                      \nonumber \\
                          &+&2\beta(y_R-1)]\ln(1-2\beta)\}^{-1},
\end{eqnarray}
where $\beta \equiv M / R$ is the dimensionless compactness parameter and $y_R \equiv y(R)$
is the solution of the following first-order differential equation (ODE):
\begin{equation}\label{Eq.10}
\frac{dy(r)}{dr}=-\frac{y(r)^2}{r}-\frac{y(r)}{r}F(r)-rQ(r),
\end{equation}
with
\begin{eqnarray}
F(r) &=&\left\{1-4\pi r^2[\varepsilon(r)-p(r)]\right\}\left[1-\frac{2m(r)}{r}\right]^{-1},        \label{Eq.11} \\
Q(r) &= &4\pi\left[5\varepsilon(r)+9p(r)+\frac{\varepsilon(r)+p(r)}{c_s^2(r)} -\frac{6}{r^2}\right]  \nonumber \\
&\times& \left[1-\frac{2m(r)}{r}\right]^{-1}  - \frac{4m^2(r)}{r^4}\left[1+\frac{4\pi{r^3}p(r)}{m(r)}\right]^2 \nonumber \\
&\times& \left[1-\frac{2m(r)}{r}\right]^{-2}, \label{Eq.12}
\end{eqnarray}
where $c_s^2(r)\equiv dp(r)/d\varepsilon(r)$ is the squared speed of sound. Starting at the center of the star,
for a given EOS, Equation~(\ref{Eq.10}) needs to be integrated self-consistently together with Equations~(\ref{Eq.5}) and (\ref{Eq.6}).
Imposing the additional boundary condition for $y$ at $r=0$ such that, $y(0)=2$,
the Love number $k_2$ and the tidal deformability $\lambda$ can be readily calculated.
One can also compute the dimensionless tidal deformability $\Lambda$, which is related to
the compactness parameter $\beta$ and the Love number $k_2$ through
\begin{equation}
\Lambda = \frac{2}{3}\frac{k_2}{\beta^5}. \label{Eq.13}
\end{equation}
The range of $\Lambda$ as a function of the stellar mass is shown in the right window of Figure~\ref{fig3}.

The total tidal effect of two neutron stars in an inspiraling binary system is given by the
mass-weighted (dimensionless) tidal deformability (see, e.g., Refs.~\cite{Hinderer:2009ca,Damour:2009vw}):
\begin{equation}
\tilde{\Lambda} = \frac{16}{13}\frac{(M_1+12M_2)M_1^4\Lambda_1+(M_2+12M_1)M_2^4\Lambda_2}{(M_1+M_2)^5}, \label{Eq.14}
\end{equation}
where $\Lambda_1=\Lambda_1(M_1)$ and $\Lambda_2=\Lambda_2(M_2)$ are the (dimensionless) tidal
deformabilities of the individual binary components. As pointed out previously~\cite{Hinderer:2009ca},
although $\Lambda$ is calculated for single neutron stars, the universality of the neutron star EOS
allows us to predict the tidal phase contribution for a given binary system from each EOS. For equal-mass
binary systems, $\tilde{\Lambda}$ reduces to $\Lambda$.  The weighted (dimensionless) deformability
$\tilde{\Lambda}$ is usually plotted as a function of the chirp mass  $\mathcal{M}=(M_1 M_2)^{3/5}/M_T^{1/5}$
for various values of the asymmetric mass ratio  $\eta=M_1M_2/M_T^2$, where $M_T=M_1+M_2$ is the total mass of
the binary.

\section{Deep Neural Networks (DNNs)}\label{sec4}

In this section, we briefly discuss the basic setup, structure, and workflow associated with implementing DNNs
for our specific application. For more extensive discussions, the reader is referred to a number of machine
learning articles \cite{LeCun2015,Emmert-Streib2020} and textbooks \cite{Goodfellow2016,Neilsen2015}.

\begin{figure}[t!]
\centering
\includegraphics[scale=0.65]{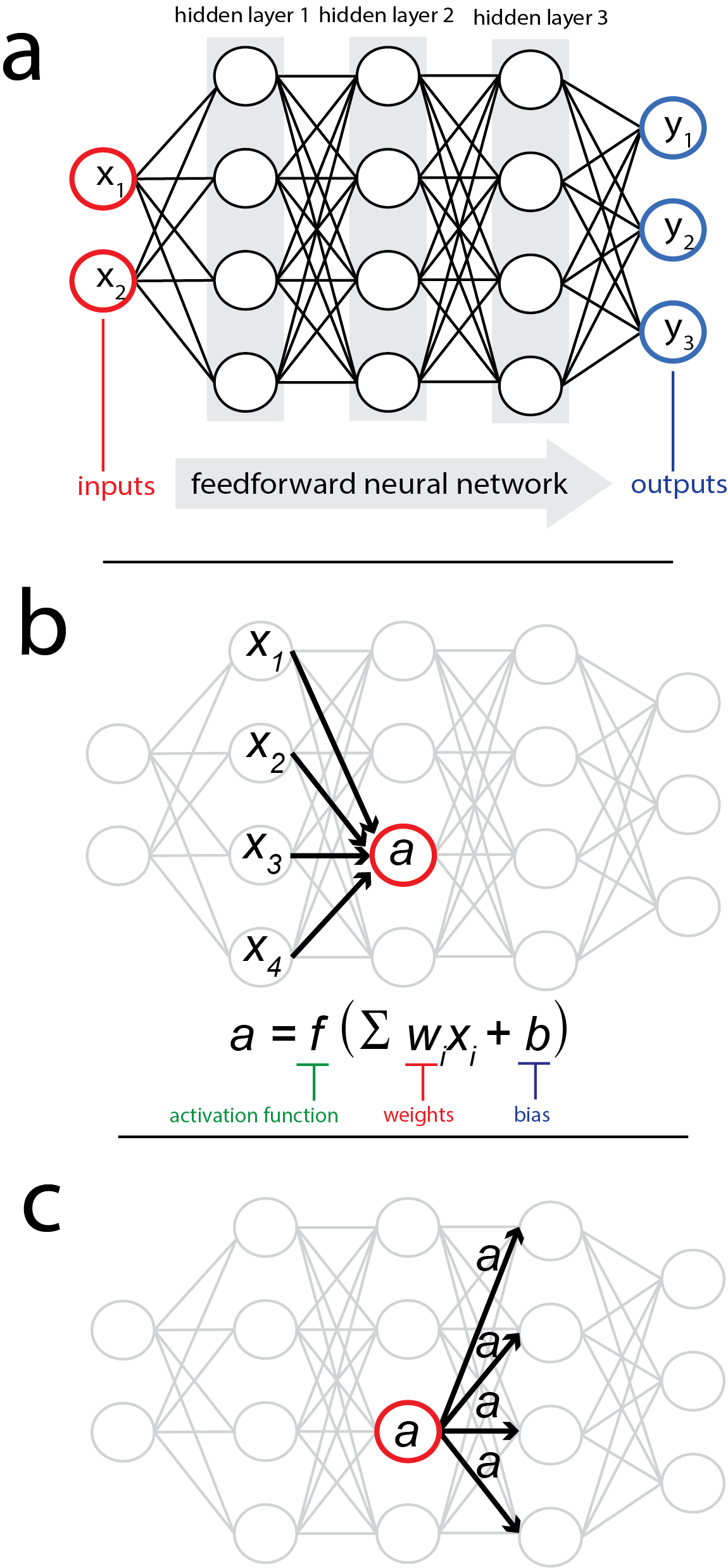}
\caption{{(\textbf{a})} Structure of an example deep neural network with input, hidden, and output layers highlighted. 
In the feedforward neural networks used in this study, calculations move from left to right, as illustrated in the
figure. {(\textbf{b})} Inputs to an example neuron from the previous layer. Also shown is the calculation
performed by the example neuron, with inputs weighted relative to one another, bias added, and an activation
function applied, in order to calculate a value $a$, referred to as the activation of the neuron. {(\textbf{c})} The activation of the
example neuron serves as one of the inputs to the next layer of neurons.  Each neuron in the successive layers of the
DNN is performing this same operation with different values of the tunable parameters $w_i$ and $b$, within the 
backpropagation algorithm. See text for details.}\label{fig4}
\end{figure}

Deep neural networks consist of processing units, named neurons, which are arranged in one to several layers (Figure~\ref{fig4}a). 
A neuron acts as a filter, performing a linear operation between the neurons in the previous layer and 
the weights associated with the neuron. A DNN typically has an input layer, followed by one or more hidden layers, and a 
final layer with one or more output neurons. As illustrated in Figure~\ref{fig4}a, in a \textit{feedforward} DNN, calculations progress 
(from left to right) starting at the input layer and moving successively through the hidden layers until reaching the output layer. 
In classification problems, the output neurons give the probabilities that an input sample belongs to a specific class. In regression 
problems, the output layer returns estimates of one or several target parameters. Each neuron in a DNN performs a simple linear 
operation. Namely, for input values $x_i$ from the previous layer, it outputs a single value $a=f ( \sum x_i w_i + b )$; see 
\mbox{Figure \ref{fig4}b,c}. {\it Activation} functions $f$ depend on the specific application but are typically chosen to be 
nonlinear or piecewise functions, such as the sigmoid or hyperbolic tangent functions \cite{LeCun2012} (Figure~\ref{fig4}b). The weights $w_i$ 
and bias $b$ are unique to each neuron and are parameters that are tuned iteratively via a backpropagation algorithm during 
the DNN training (Figure~\ref{fig4}b). Neuron $k$ in layer $m$ accepts the outputs $a_1^{m-1}$, $a_2^{m-1}$,\ldots , $a_N^{m-1}$ from 
all $N$ from the previous layer $m-1$ and computes a single value $a_k^m=f ( \sum a^{m-1}_i w_i + b )$ that is broadcasted to 
every neuron in layer $m+1$ \cite{Neilsen2015}.

In this analysis, we apply a DL approach and formulate a regression problem, where the inputs to the DNNs consist of $M(R)$ or 
$M(\Lambda)$ sequences, while the outputs consist of $E_{sym}(\rho)$ estimates. Accordingly, the data sets consist of 
$E_{sym}(\rho)$ samples and $M(R)$, or $M(\Lambda)$, sequences. We use the EOS metamodel discussed in Section \ref{sec2},
with the SNM part of the EOS kept fixed, and vary only the symmetry energy. In particular, we set $E_0 = 15.9$~MeV, 
$K_0 = 240$ MeV, $J_0 = 0$ MeV, $S_0 = 31.7$ MeV, $J_{sym} = 0$, and vary only $L$ and $K_{sym}$ in Equations~(\ref{Eq.2})
and (\ref{Eq.3}). Specifically, the values of $L$ and $K_{sym}$ are sampled randomly from their respective ranges of 
[30.6--86.8] MeV and [$-$400--100] MeV. {Recently, the latest results of the PREX collaboration suggested a rather
high value of $L$ with an upper limit at 143~MeV~\cite{PREX:2021umo}. Examining the effect of higher $L$ values is left to following works.}
The resultant EOSs $p(\varepsilon)$ are checked regarding whether they satisfy (i) the microscopic
stability condition, i.e., $\frac{dp}{d\varepsilon}\geq 0$, and (ii) the causality condition, i.e., the speed of sound 
$c_s\equiv \sqrt{\frac{dp}{d\varepsilon}}\geq c$, which restricts the values of $L$ and $K_{sym}$ and the
$E_{sym}(\rho)$ samples. For each EOS, the NS structure equations are solved to obtain $M(R)$ and $M(\Lambda)$
sequences. To simulate NS observational data, from a given genuine $M-R$ (or $M-\Lambda$) sequence, 
we randomly choose 50 points in the range of 1M$_{\odot}$ to 2M$_{\odot}$ \endnote{At present, such a large 
number of simultaneous NS mass and radius (or $M$ and $\Lambda$) measurements may look too optimistic. However, with
the rapid advent of the next generation of telescopes and GW detectors, a much greater number of NS observations
is expected in the near future.}. Then, each sample input is a vector of dimension 100, with the two arrays of $M$ and 
$R$ (or $M$ and $\Lambda$) values concatenated. Similarly, each output sample is a vector of dimension 100 representing
an estimated $E_{sym}(\rho)$ in the density range of $\sim$$0.5 \rho/\rho_0$ to $5\rho/\rho_0$. In this respect, the DNN maps 
an input $M(R)$ or $M(\Lambda)$ sequence to an output $E_{sym}(\rho)$. Realistic NS observations inevitably accrue errors, 
which result in corresponding uncertainties when reconstructing the symmetry energy, and/or the EOS. For the purpose of this
analysis, however, we do not take into account NS observational errors and uncertainties. This work should be regarded as a
\textit{proof-of-concept study}, and the application to realistic NS data is left to a future article.

\begin{figure*}[t!]
\centering
\includegraphics[scale=0.5]{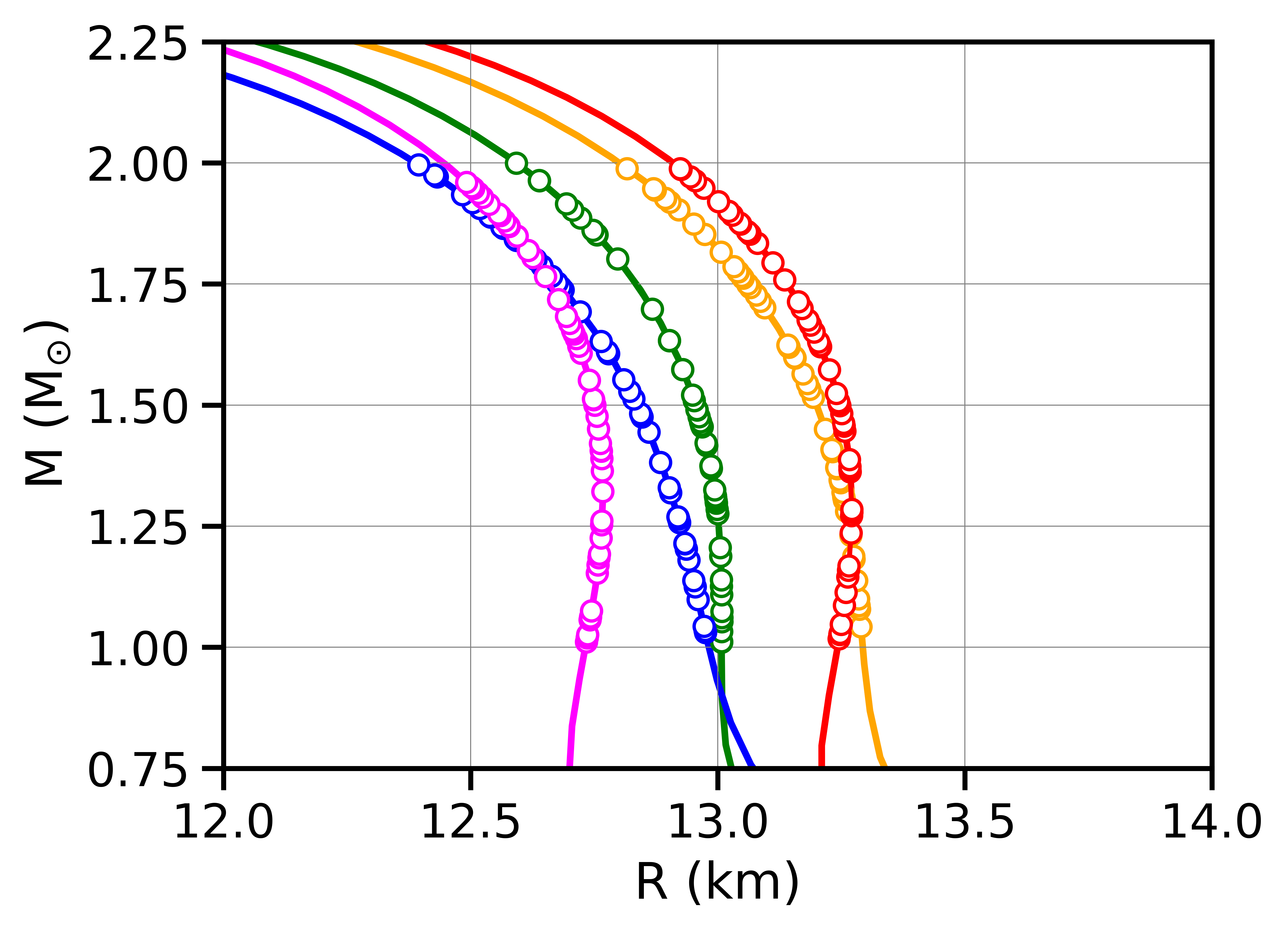}
\includegraphics[scale=0.5]{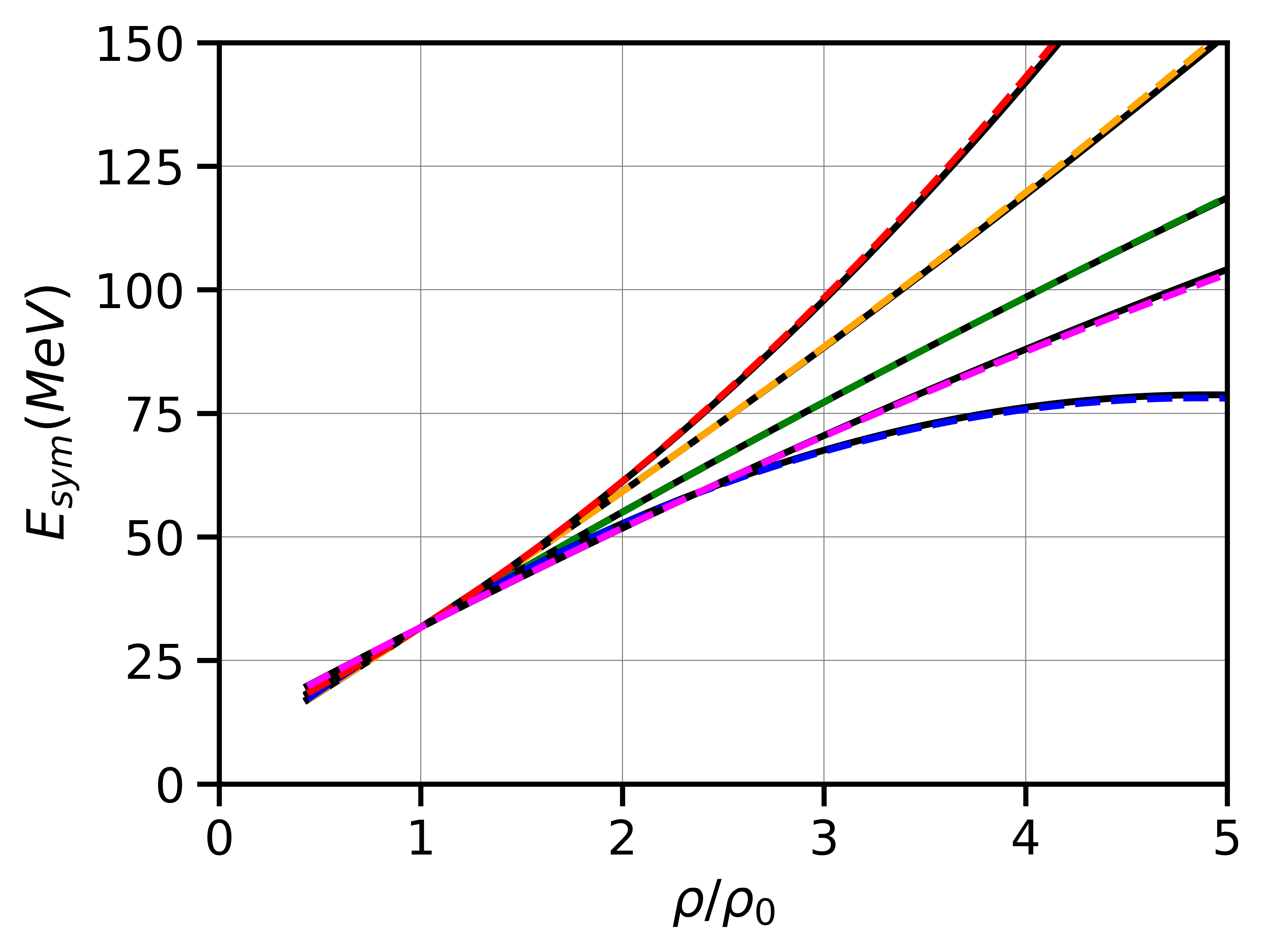}
\caption{Example input $M(R)$ sequences (\textbf{left window}) and corresponding estimated $E_{sym}(\rho)$ (\textbf{right window}).
The input samples consist of 50 randomly selected points, denoted by the ``o'' characters, from the genuine $M(R)$ curves,
denoted by the solid lines, in the range of 1--2 M$_{\odot}$. The output data samples consist of 100 $E_{sym}(\rho)$ points 
in the range of $\sim$0.4--5~$\rho_0$. Broken colored lines in the right window denote the estimated $E_{sym}(\rho)$ and 
the solid lines represent the exact $E_{sym}(\rho)$. Same curve colors in both windows denote pairs of input
$M(R)$ sequences and corresponding output symmetry energy.}\label{fig5}
\end{figure*}

In supervised learning, the data sets are divided into training, validation, and testing data. The training data set is used by
the DNN to learn from, the validation data are used to verify whether the network is learning correctly, and the testing data
are used to assess the performance of the trained model. The training data sets used in this work consist of 40,000 independent
$M(R)$, or $M(\Lambda)$, sequences representing the DNN inputs, and 40,000 matching $E_{sym}(\rho)$ samples representing 
the DNN outputs. The validation and testing data sets consist of $\sim$1000 input samples and the same number of output samples each.

The neural networks used here are feedforward DNNs with 10 hidden, dense, fully connected layers of dimension 100, and 
{\it ReLU} activation functions. The first layer has a linear activation function and corresponds to the input to the neural network, 
which, in this case, is a one-dimensional concatenated vector containing the NS $M$ and $R$ (or $\Lambda$) values for a given
$M(R)$ (or $M(\Lambda)$) sequence. At the end, there is a linear output layer of dimension 100 returning the estimated 
$E_{sym}(\rho)$. The network design was optimized by fine-tuning multiple hyper-parameters, which include here the number 
and type of network layers, the number of neurons in each layer, and the type of activation function. The optimal network architecture 
was determined through multiple experiments and tuning of the hyper-parameters. The feedforward DNN used in this work and its
functionality is shown schematically in Figure~\ref{fig4}.

To build and train the neural networks, we used the Python toolkit Keras (\url{https://keras.io} (accessed on September 28, 2021)), which provides a high-level application 
programming interface (API) to the TensorFlow \cite{TF} (\url{https://www.tensorflow.org}  (accessed on September 28, 2021)) deep learning library. We applied 
the technique of stochastic gradient descent with an adaptive learning rate with the ADAM method \cite{Adam} with the AMSgrad modification \cite{Adam2}. 
To train the  DNNs, we used an initial learning rate of 0.003 and chose a batch size of 500. During each training session, the number of epochs was 
limited to 2000, or until the validation error stopped decreasing. The training of the DNNs was performed on an NVIDIA Tesla V100 GPU and the 
size of the mini-batches was chosen automatically depending on the specifics of the GPU and data sets.  We used the mean squared 
error (MSE) as a cost (or loss) function.

\section{Results}\label{sec5}

We first examine the ability of the DNN to reconstruct the $E_{sym}(\rho)$ from a set of mass and radius $M-R$ measurements that
may result from electromagnetic observations of neutron stars, such as those from the NICER mission, for instance. Specifically,
we apply the trained DNN, described in the previous section, to a test data set containing $\sim$1000 simulated $M(R)$ sequences, and
compare the corresponding estimated output $E_{sym}(\rho)$ with the exact symmetry energy for each sample. In Figure~\ref{fig5}, we show
results for five representative examples from the test data set. It is seen that the estimated symmetry energy (broken colored lines)
for each input $M-R$ sequence matches almost exactly the ``true'' $E_{sym}(\rho)$ (solid black lines) over the entire density range considered
here. The results are very similar for the rest of the test data samples. Quantitatively, at $5\rho_0$, the mean absolute error over the whole test 
data set is 1.2 MeV, with a standard deviation of 0.8 MeV, where the errors are even smaller at lower densities. Choosing different ensembles
of randomly selected points from the genuine $M(R)$ curves does not alter appreciably the accuracy with which the symmetry energy
is estimated. 

\begin{figure*}[t!]
\centering
\includegraphics[scale=0.5]{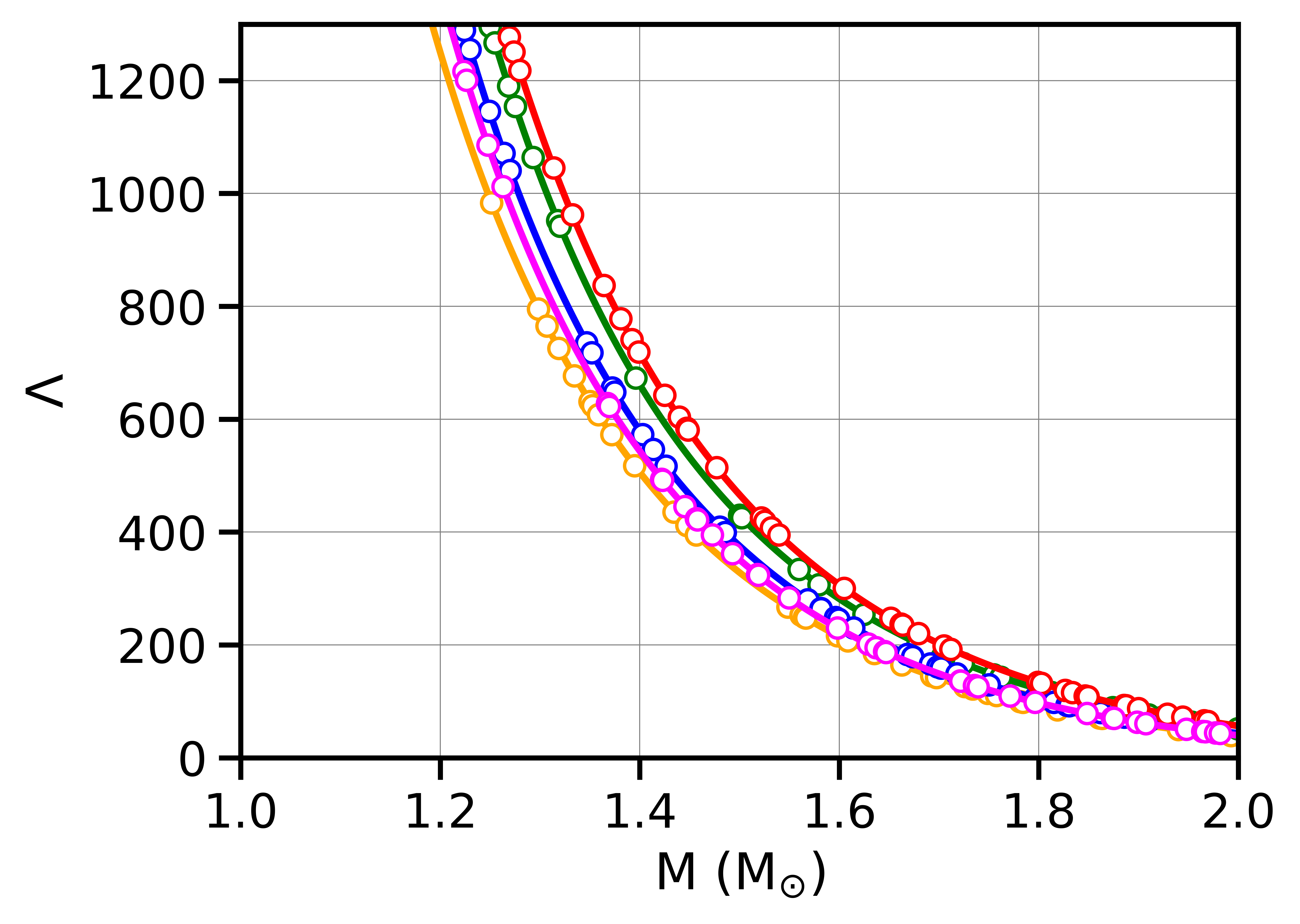}
\includegraphics[scale=0.5]{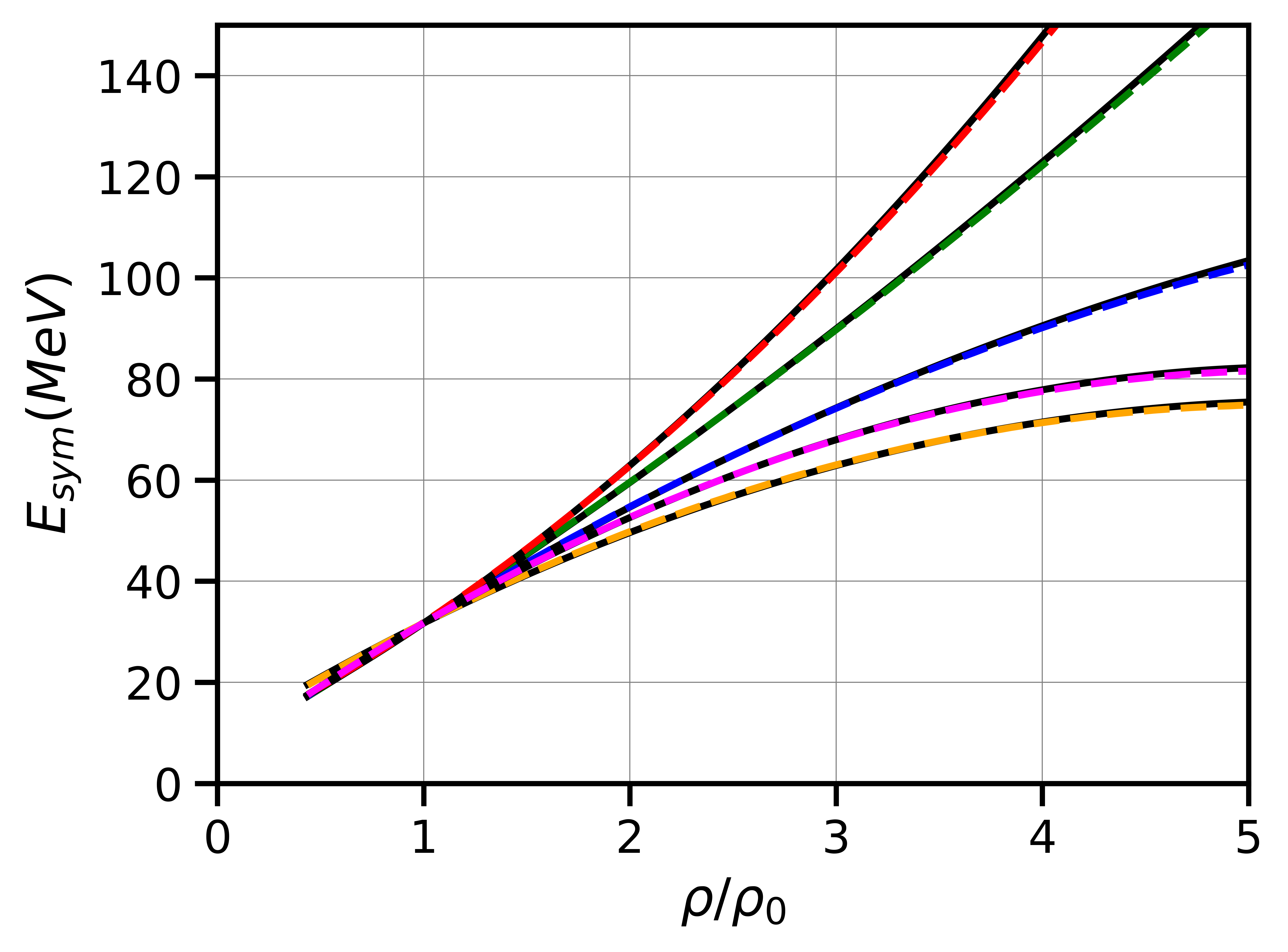}
\caption{Example input $M(\Lambda)$ sequences (\textbf{left window}) and corresponding estimated $E_{sym}(\rho)$ (\textbf{right~window}).
The input samples consist of 50 randomly selected points, denoted by the ``o'' characters, from the genuine $M(\Lambda)$ curves,
denoted by the solid lines, in the range of 1--2 M$_{\odot}$. All other figure features are the same as in Figure~\ref{fig5}. See text for
details.}\label{fig6}
\end{figure*}

At this point, we need to reiterate that realistic NS observations inevitably carry uncertainties, which would result in
corresponding uncertainties in the estimated symmetry energy. However, this work should be considered a ``proof-of-concept study''
and realistic applications are left for following articles. In addition, these results are based on the assumption that we have 50 NS $M-R$
observations. Although, at present, such a large number may seem rather optimistic, with the advent of the next generation of
electromagnetic observatories, this circumstance is rapidly changing as many more NS electromagnetic observations are expected in the 
future. These results clearly demonstrate the ability of the DL-based approach to extract the nuclear symmetry energy accurately
from NS mass and radius measurements, given that observational data of sufficient quality exist.

We next look at the ability of the neural network to extract the symmetry energy entirely from GW data from compact
binary mergers involving neutron stars, particularly BNS events. Specifically, we assume a set of mass and tidal deformability
$M - \Lambda$ measurements that may result from the LIGO/VIRGO/KAGRA GW detectors, and from the next-generation GW 
ground-based (e.g., Einstein Telescope and the Cosmic Explorer) and space-borne (e.g., LISA) observatories, in the future. 
In this case, the trained DNN model is applied to a test data set containing $\sim$1000 simulated $M(\Lambda)$ sequences,
and the estimated output $E_{sym}(\rho)$ is compared with the exact symmetry energy for each data sample. The results are
summarized in Figure~\ref{fig6}, which shows five representative examples from the test data set. Again, the estimated $E_{sym}(\rho)$
(broken colored lines) follows very closely the ``true'' symmetry energy (solid black lines) for each input $M - \Lambda$ sequence
over the entire density range, and the results follow a very similar trend for all samples in the test data set. The overall performance 
of the DNN is again assessed by looking at the mean absolute error, which, at $5\rho_0$, is 0.8 MeV, with a standard deviation of 
0.4 MeV. As discussed in the previous section, the input of the neural network consists of 50 randomly chosen points from a given 
genuine $M(\Lambda)$ curve in the range of 1--2 M$_{\odot}$, and the output consists of 100 fixed points of $E_{sym}(\rho)$ in the
range $\sim$0.4--5 $\rho_0$. Using different input ensembles of randomly selected points does not change significantly the results
presented in Figure~\ref{fig6}. These findings show that the $E_{sym}(\rho)$ can be extracted accurately, via artificial neural networks, 
directly from GW observational data of neutron stars. They also emphasize the importance of the DL-based approach for extracting the
symmetry energy, the most uncertain component of the dense matter EOS, from GW observations. As in the case of NS mass and radius
measurements, GW observations of neutron stars inevitably accrue errors, which result in corresponding uncertainties in the extracted
symmetry energy, and the EOS. As already mentioned, realistic applications of the DL method  are left to following works. 

These novel computational techniques will become particularly important in the future with the advent of the next generation 
GW observatories, when millions of GW events will be routinely detected per year, with at least several events involving neutron 
stars per day. As more such GW events are detected and characterized, this data-driven approach will eventually allow us to map precisely the
GW observations of neutron stars to the $E_{sym}(\rho)$, and in turn, the underlying EOS of dense nuclear matter. 

\section{Summary and Outlook}\label{sec6}

We have demonstrated, for the first time, the reconstruction of the nuclear symmetry energy directly from MMA observations of neutron stars using
deep learning approaches. Specifically, we have shown that deep neural networks can extract $E_{sym}(\rho)$ accurately from either
mass and radius $M - R$ or mass and tidal deformability $M - \Lambda$ astrophysical measurements of neutron stars. These results are a step
towards realizing the goal of determining the EOS of dense nuclear matter, and they underline the importance and potential of the DL-based
approach in the era of multi-messenger astrophysics, where an ever-increasing volume of MMA data is becoming rapidly available.

Future directions include considering the complete set of EOS parameters in Equations~(\ref{Eq.2}) and (\ref{Eq.3}). Specifically, we will also take into account
the effect of $J_0$ and $J_{sym}$, which are set to $J_0=J_{sym}=0$ in the current analysis. Despite the large current uncertainties of these higher-order
parameters, their inclusion will allow us to model a much wider range of realistic EOSs, and thus enable a direct comparison of $E_{sym}(\rho)$
estimates, obtained with the DL techniques developed in this work, with realistic symmetry energies. Astrophysical observations of neutron stars 
inevitably carry error uncertainties, which lead to corresponding errors and uncertainties in the extracted symmetry energy, and EOS. For realistic
applications of our approach, the effect of empirical errors and uncertainties needs to be considered and implemented consistently in the formalism.
This can be achieved, for instance, by recasting the $E_{sym}(\rho)$ regression problem into a probabilistic framework. In subsequent works, we plan
to implement Bayesian neural networks to perform the symmetry energy inference task. In this paradigm, instead of having deterministic values, 
the weights of these networks are characterized by probabilistic distributions by placing a prior over the network weights \cite{Perreault2017}.
Finally, we also plan to investigate likelihood-free inference methods based on normalizing flows \cite{Kobyzev2021}. By applying a series of nonlinear transforms 
to a simple posterior shape (e.g., a multivariate Gaussian), the flow is able to reproduce complex posteriors without evaluating the likelihood directly.
These techniques have already gained considerable interest in the research community, and, recently, a likelihood-free inference method based 
on normalizing flows was applied \cite{Dax:2021tsq} to rapidly perform parameter estimation of eight BBH GW events in the first LIGO catalog, 
GWTC-1 \cite{LIGOScientific:2018mvr}. Such rapid processing will be particularly important for next-generation space telescopes and GW detectors, 
whose sensitivity goals will allow for the detection and observation of compact binary collisions, and neutron stars, throughout the history of the universe, 
exceeding a million events per year, with thousands of BNS detections/year alone. Conventional Bayesian inference approaches are not 
scalable to the study of thousands of BNS  events per year, and modern normalizing flow models could certainly help to extract BNS parameters 
promptly and accurately.

Ultimately, as more events involving neutron stars are observed, these modern data-driven approaches will allow us to rapidly process the ever-increasing
amount of neutron star observational data and determine precisely the nuclear symmetry energy and the EOS of dense nuclear matter.
\\\\
\noindent\textbf{Data Availability:} Codes and data from this analysis are available upon request from the author.

\section*{Acknowledgements}
The computations in this paper were run on the FASRC Cannon cluster supported by the FAS Division of Science Research Computing Group at Harvard University.


\begin{references}
\bibitem{NAP2011}
The National Academies Press. {\it New Worlds, New Horizons in Astronomy and Astrophysics}; The National Academies Press: Washington,
DC, USA, 2011.  Available online: https://www.nap.edu/catalog/12951/ new-worlds-new-horizons-in-astronomy-and-astrophysics.

\bibitem{NAP2012}
The National Academies Press. {\it Nuclear Physics: Exploring the Heart of Matter}; Report of the Committee on the Assessment
of and Outlook for Nuclear Physics; The National Academies Press: Washington, DC, USA, 2012. Available online: 
https://www.nap.edu/catalog/13438/nuclear-physics-exploring-the-heart-of-matter.

\bibitem{USLongRangePlan2015}
2015 U.S. Long Range Plan for Nuclear Sciences. Available online: \url{https://www.osti.gov/servlets/purl/1296778}.

\bibitem{EPJA2014}
Li, B.A.; Ramos, \`{A}.; Verde, G.; Vida\~{n}a, I. Topical Issue on Nuclear Symmetry Energy. {\it Eur. Phys. J. A} {\bf 2014}, \emph{50}, 9.
https://doi.org/10.1140/epja/i2014-14009-x.

\bibitem{LiUniverse2021}
Li, B.-A.; Cai, B.-J.; Xie, W.-J.; Zhang, N.-B. Progress in Constraining Nuclear Symmetry Energy Using Neutron Star Observables Since GW170817.
{\it Universe} {\bf 2021}, \emph{7}, 182. https://doi.org/10.3390/universe7060182.

\bibitem{Science2002}
Danielewicz, P.; Lacey, R.; Lynch, W.G. Determination of the equation of state of dense matter. {\it Science} {\bf 2002}, \emph{298}, 1592--1596.
https://doi.org/10.1126/science.1078070.

\bibitem{PhysRep2005v1}
Baran, V.; Colonna, M.; Greco, V.; Di Toro, M. Reaction dynamics with exotic nuclei. {\it Phys. Rep.} {\it 2005}, \emph{410}, 335--466.
https://doi.org/10.1016/j.physrep.2004.12.004.

\bibitem{PhysRep2005v2}
Steiner, A.W.; Prakash, M.; Lattimer, J.M.; Ellis, P.J. Isospin asymmetry in nuclei and neutron stars. {\it Phys. Rep.} {\bf 2005}, \emph{411}, 325--375.
https://doi.org/10.1016/j.physrep.2005.02.004.

\bibitem{PRC2012}
Tsang, M.B.; Stone, J.R.; Camera, F.; Danielewicz, P.; Gandolfi, S.; Hebeler, K.; Horowitz, C.J.; Lee, J.; Lynch, W.G.; Kohley, Z.; et al.
Constraints on the symmetry energy and neutron skins from experiments and theory. {\it Phys. Rev. C} {\bf 2012}, \emph{86}, 015803.
https://doi.org/10.1103/PhysRevC.86.015803.

\bibitem{PPNP2016}
Baldo, M.; Burgio, G.F. The nuclear symmetry energy. {\it Prog. Part. Nucl. Phys.} {\bf 2016}, \emph{91}, 203--258. https://doi.org/10.1016/j.ppnp.2016.06.006.

\bibitem{NPN2017}
{Li, B.-A. Nuclear symmetry energy extracted from laboratory experiments. {\it Nucl. Phys. News} {\bf 2017}, \emph{27}, 7--11.}

\bibitem{PPNP2018}
Li, B.-A.; Cai, B.J.; Chen, L.W.; Xu, J. Nucleon effective masses in neutron-rich matter. {\it Prog. Part. Nucl. Phys.} {\bf 2018}, \emph{99}, 29--119.
https://doi.org/10.1016/j.ppnp.2018.01.001.

\bibitem{BurgioVidanaUniverse2020}
Burgio, G.F.; Vida\~{n}a, I. The Equation of State of Nuclear Matter: From Finite Nuclei to Neutron Stars. {\it Universe} {\textbf{2020}}, \emph{6}, 119. https://doi.org/10.3390/universe6080119.

\bibitem{Lattimer2001}
Lattimer, J.M.; Prakash, M. Neutron star structure and the equation of state. {\it Astrophys. J.} {\bf 2001}, \emph{550}, 426--442. https://doi.org/10.1086/319702.

\bibitem{Lattimer2016}
Lattimer, J.M.; Prakash, M. The equation of state of hot, dense matter and neutron stars. {\it Phys. Rep.} {\bf 2016}, \emph{621}, 127--164. https://doi.org/10.1016/j.physrep.2015.12.005.

\bibitem{Watts2016}
Watts, A.L.; Andersson, N.; Chakrabarty, D.; Feroci, M.; Hebeler, K.; Israel, G.; Lamb, F.K.; Miller, M.C.; Morsink, S.; \"{O}zel, F.; et al.
Colloquium: Measuring the neutron star equation of state using X-ray timing. {\it Rev. Mod. Phys.} {\bf 2016}, \emph{88}, 021001. https://doi.org/10.1103/RevModPhys.88.021001.

\bibitem{Ozel2016}
\"{O}zel, F.; Freire, P. Masses, radii, and the equation of state of neutron stars. {\it Annu. Rev. Astron. Astrophys.} {\bf 2016}, \emph{88}, 401--440. https://doi.org/10.1146/annurev-astro-081915-023322.

\bibitem{Oertel2017}
Oertel, M.; Hempel, M.; Kl\"{a}hn, T.; Typel, S. Equations of state for supernovae and compact stars. {\it Rev. Mod. Phys.} {\bf 2017}, \emph{89}, 015007. https://doi.org/10.1103/RevModPhys.89.015007.

\bibitem{Baiotti2019}
Baiotti, L. Gravitational waves from neutron star mergers and their relation to the nuclear equation of state. {\it Prog. Part. Nucl. Phys.}
{\textbf{2019}}, \emph{109}, 103714.
https://doi.org/10.1016/j.ppnp.2019.103714.

\bibitem{EPJA2019}
Li, B.-A.; Krastev, P.G.; Wen, D.H.; Zhang, N.B. Towards understanding astrophysical effects of nuclear symmetry energy. 
{\it Eur. Phys. J. A} {\bf 2019}, \emph{55}, 117. 
https://doi.org/10.1140/epja/i2019-12780-8.

\bibitem{Weber2007}
Weber, F.; Negreiros, R.; Roseneld, P.; Stejner, M. Pulsars as astrophysical laboratories for nuclear and particle physics. {\it Prog. Part. Nucl. Phys.} 
{\bf 2007}, \emph{59}, 94--113. 
https://doi.org/10.1016/j.ppnp.2006.12.008.

\bibitem{Alford2019}
Alford, M.G.; Han, S.; Schwenzer, K. Signatures for quark matter from multi-messenger observations. {\it J. Phys. G Nucl. Part. Phys.}
{\bf 2019}, \emph{46}, 114001.
https://doi.org/10.1088/1361-6471/ab337a.

\bibitem{Capano2020}
Capano, C.D.; Tews, I.; Brown, S.M.; Margalit, B.; De, S.; Kumar, S.; Brown, D.A.; Krishnan, B.; Reddy, S. Stringent constraints on
neutron-star radii from multimessenger observations and nuclear theory. {\it Nat. Astron.} {\bf 2020}, \emph{4}, 625--632.
https://doi.org/10.1038/s41550-020-1014-6.

\bibitem{Blaschke2020}
Blaschke, D.; Ayriyan, A.; Alvarez-Castillo, D.E.; Grigorian, H.~Was GW170817 a canonical neutron star merger? Bayesian analysis
with a third family of compact stars. {\it Universe} {\bf 2020}, \emph{6}, 81.
https://doi.org/10.3390/universe6060081.

\bibitem{Chatziioannou2020}
Chatziioannou, K. Neutron-star tidal deformability and equation-of-state constraints. {\it Gen. Relativ. Gravit.} {\bf 2020}, \emph{52}, 109.
https://doi.org/10.1007/s10714-020-02754-3.

\bibitem{Annala2018}
Annala, E.; Gorda, T.; Kurkela, A.; Vuorinen, A. Gravitational-Wave Constraints on the Neutron-Star-Matter Equation of State.
{\it Phys. Rev. Lett.} {\bf 2018}, \emph{120}, 172703.
https://doi.org/10.1103/PhysRevLett.120.172703.

\bibitem{Kievsky2018}
Kievsky, A.; Viviani, M.; Logoteta, D.; Bombaci, I.; Girlanda, L. Correlations imposed by the unitary limit between few-nucleon
systems and compact stellar systems. {\it Phys. Rev. Lett.} {\bf 2018}, \emph{121}, 072901.
https://doi.org/10.1103/PhysRevLett.121.072701.

\bibitem{Landry2020}
Landry, P.; Essick, R.; Chatziioannou, K. Nonparametric constraints on neutron star matter with existing and upcoming
gravitational wave and pulsar observations. {\it Phys. Rev. D} {\bf 2020}, \emph{101}, 123007.
https://doi.org/10.1103/PhysRevD.101.123007.

\bibitem{Dietrich2020}
Dietrich, T.; Coughlin, M.W.; Pang, P.T.H.; Bulla, M.; Heinzel, J.; Issa, L.; Tews, I.; Antier, S. Multimessenger constraints on the
neutron-star equation of state and the Hubble constant. {\it Science} {\bf 2020}, \emph{370}, 1450--1453.
http://dx.doi.org/10.1126/science.abb4317.

\bibitem{Stone2021}
Stone, J.R. Nuclear Physics and Astrophysics Constraints on the High Density Matter Equation of State. {\it Universe} {\bf 2021}, \emph{7}, 257. 
https://doi.org/10.3390/universe7080257.

\bibitem{Li2020}
Li, A.; Zhu, Z.Y.; Zhou, E.P.; Dong, J.M.; Hu, J.N.; Xia, C.J. Neutron star equation of state: Quark mean-field (QMF) modeling and
applications. {\it J. High Energy Astrophys.} {\bf 2020}, \emph{28}, 19--46.
https://doi.org/10.1016/j.jheap.2020.07.001.

\bibitem{Burgio:2021vgk}
Burgio, G.F.; Vida\~{n}a I.; Schulze, H.-J.; Wei, J.-B. Neutron stars and the nuclear equation of state. {\it Prog. Part. Nucl. Phys.} {\bf 2021}, \emph{120}, 103879.
https://doi.org/10.1016/j.ppnp.2021.103879.

\bibitem{Burgio:2021bzy}
Burgio, G.F.; Schulze, H.J.; Vidaña, I.; Wei, J.B. A Modern View of the Equation of State in Nuclear and Neutron Star Matter. {\it Symmetry} {\bf 2021}, 13, 400. 
https://doi.org/10.3390/sym13030400

\bibitem{KrastevJPG2019}
Krastev, P.G.; Li, B.-A. Imprints of the nuclear symmetry energy on the tidal deformability of neutron stars. {\it J. Phys. G} {\bf 2019}, \emph{46}, 074001.
https://doi.org/10.1088/1361-6471/ab1a7a.

\bibitem{Raithel:2019ejc}
Raithel, C.A.; \"{O}zel, F. Measurement of the nuclear symmetry energy parameters from gravitational wave events. {\it Astrophys. J.} {\bf 2019}, \emph{885}, 121.
https://doi.org/10.3847/1538-4357/ab48e6.

\bibitem{NICER2017}
Gendreau, K.; Arzoumanian, Z. Searching for a pulse. {\it Nat. Astron.} {\bf 2017}, \emph{1}, 895. https://doi.org/10.1038/s41550-017-0301-3.

\bibitem{aLIGO2015}
Aasi, J.; Abbott, B.P.; Abbott, R.; Abbott, T.; Abernathy, M.R.; Ackley, K.; Adams, C.; Adams, T.; Addesso, P.; Adhikari, R.X.; et~al. Advanced LIGO. {\it Class. Quant. Grav.} {\bf 2015}, \emph{32}, 074001. 
https://doi.org/10.1088/0264-9381/32/7/074001.

\bibitem{VIRGO:2014yos}
Acernese, F.A.; Agathos, M.; Agatsuma, K.; Aisa, D.; Allemandou, N.; Allocca, A.; Amarni, J.; Astone, P.; Balestri, G.; Ballardin, G.; et~al. Advanced Virgo{:} a second-generation interferometric gravitational wave detector. {\it Class. Quant. Grav.} {\bf 2015}, \emph{32}, 024001. 
https://doi.org/10.1088/0264-9381/32/2/024001.

\bibitem{KAGRA2019}
KAGRA Collaboration. KAGRA: 2.5 generation interferometric gravitational wave detector. {\it Nat. Astron.} {\bf 2019}, \emph{3}, 35-40. 
https://doi.org/10.1038/s41550-018-0658-y.

\bibitem{Bombaci1991}
Bombaci, I.; Lombardo, U. Asymmetric nuclear matter equation of state. {\it Phys. Rev. C} {\bf 1991}, \emph{44}, 1892. 

\bibitem{Hebeler2010}
Hebeler, K.; Schwenk, A. Chiral three-nucleon forces and neutron matter. {\it Phys. Rev. C} {\bf 2010} \emph{82}, 014314.

\bibitem{Tews2013}
Tews, I.; Kr\"{u}ger, T.; Hebeler, K.; Schwenk, A. Neutron matter at next-to-next-to-next-to-leading order in chiral effective field theory. {\it Phys. Rev. Lett.}
{\bf 2013}, \emph{110}, 032504. https://doi.org/10.1103/PhysRevLett.110.032504

\bibitem{Holt2013}
Holt, J.W.; Kaiser, N.; Weise, W. Nuclear chiral dynamics and thermodynamics. {\it Prog. Part. Nucl. Phys.} {\bf 2013}, \emph{73}, 35. https://doi.org/10.1016/j.ppnp.2013.08.001

\bibitem{Hagen2014}
Hagen, G.; Papenbrock, T.; Ekstr\"{o}m, A.; Wendt, K.A.; Baardsen, G.; Gandolfi, S.; Hjorth-Jensen, M.; Horowitz, C.J. Coupled-cluster calculations of nucleonic matter. {\it Phys. Rev. C} {\bf 2014}, \emph{89}, 014319.
https://doi.org/10.1103/PhysRevC.89.014319.

\bibitem{Roggero2014}
Roggero, A.; Mukherjee, A.; Pederiva, F. Quantum Monte Carlo calculations of neutron matter with non-local chiral interactions. 
{\it Phys. Rev. Lett.} {\bf 2014}, \emph{112}, 221103.
https://doi.org/10.1103/PhysRevLett.112.221103.

\bibitem{Machleidt2011}
Machleidt, R.; Entem, D.R. Chiral effective field theory and nuclear forces. {\it Phys. Rep.} {\bf 2011}, \emph{503}, 1--75. 
\bibitem{Wlazlowski2014}
Wlaz{\l}owski, G.; Holt, J.W.; Moroz, S.; Bulgac, A.; Roche, K.J. Auxiliary-Field Quantum Monte Carlo Simulations of Neutron 
Matter in Chiral Effective Field Theory. {\it Phys. Rev. Lett.} {\bf 2014}, \emph{113}, 182503.
https://doi.org/10.1103/PhysRevLett.113.182503.

\bibitem{Tews2018}
Tews, I.; Carlson, J.; Gandolfi, S.; Reddy, S. Constraining the speed of sound inside neutron stars with chiral effective field theory 
interactions and observations. {\it Astrophys. J.} {\bf 2018}, \emph{860}, 149.
https://doi.org/10.3847/1538-4357/aac267.

\bibitem{Drischler2020}
Drischler, C.; Furnstahl, R.J.; Melendez, J.A.; Phillips, D.R. How Well Do We Know the Neutron-Matter Equation of State at the Densities 
Inside Neutron Stars? A Bayesian Approach with Correlated Uncertainties. {\it Phys. Rev. Lett.} {\bf 2020}, \emph{125}, 202702.
https://doi.org/10.1103/PhysRevLett.125.202702.

\bibitem{Drischler2021}
Drischler, C.; Holt, J.W.; Wellenhofer, C. Chiral Effective Field Theory and the High-Density Nuclear Equation of State. {\it Ann. Rev. Nucl. Part. Sci.} {\bf 2021}, \emph{71}, 1. https://doi.org/10.1146/annurev-nucl-102419-041903

\bibitem{Freedman1977v1}
Freedman, B.A.; McLerran,L.D. Fermions and Gauge Vector Mesons at Finite Temperature and Density. 1. Formal Techniques. {\it Phys. Rev. D} {\bf 1977}, \emph{16}, 1130.
https://doi.org/10.1103/PhysRevD.16.1130

\bibitem{Freedman1977v2}
Freedman, B.A.; McLerran, L.D. Fermions and Gauge Vector Mesons at Finite Temperature and Density. 3. The Ground State 
Energy of a Relativistic Quark Gas. {\it Phys. Rev. D} {\bf 1977}, \emph{16}, 1169.
https://doi.org/10.1103/PhysRevD.16.1169.

\bibitem{Baluni1978}
Baluni, V. Nonabelian Gauge Theories of Fermi Systems: Chromotheory of Highly Condensed Matter. {\it Phys. Rev. D} {\bf 1978}, \emph{17}, 2092.
https://doi.org/10.1103/PhysRevD.17.2092.

\bibitem{Kurkela2010}
Kurkela, A.; Romatschke, P.; Vuorinen, A. Cold Quark Matter. {\it Phys. Rev. D} {\bf 2010}, 81, 105021. 

\bibitem{Fraga2014}
Fraga, E.S.; Kurkela, A.; Vuorinen, A. Interacting quark matter equation of state for compact stars. {\it Astrophys. J. Lett.} {\bf 2014}, \emph{781}, L25.
https://doi.org/10.1088/2041-8205/781/2/L25.

\bibitem{Gorda2018}
Gorda, T.; Kurkela, A.; Romatschke, P.; S\"{a}ppi, M.; Vuorinen, A. Next-to-Next-to-Next-to-Leading Order Pressure of Cold Quark Matter: Leading Logarithm.
{\it Phys. Rev. Lett.} {\bf 2018}, \emph{121}, 202701.
https://doi.org/10.1103/PhysRevLett.121.202701.

\bibitem{Ghiglieri2020}
Ghiglieri, J.; Kurkela, A.; Strickland, M.; Vuorinen, A. Perturbative Thermal QCD: Formalism and Applications. {\it Phys. Rept.} {\bf 2020}, \emph{880}, 1.
https://doi.org/10.1016/j.physrep.2020.07.004.

\bibitem{Fujimoto:2021zas}
Fujimoto, Y.; Fukushima, K.; Murase, K. Extensive Studies of the Neutron Star Equation of State from the Deep Learning Inference with the Observational Data Augmentation. {\it J. High Energ. Phys.} {\bf 2021}, \emph{3}, 273.
https://doi.org/10.1007/JHEP03(2021)273.

\bibitem{Aarts2016}
Aarts, G. Introductory lectures on lattice QCD at nonzero baryon number. {\it J. Phys. Conf. Ser.} {\bf 2016}, \emph{706}, 022004.
https://doi.org/10.1088/1742-6596/706/2/022004.

\bibitem{LiPhysRep2008}
Li, B.-A.; Chen, L.-W.; Ko,C.M. Recent progress and new challenges in isospin physics with heavy-ion reactions. {\it Phys. Rep.} {\bf 2008}, \emph{464}, 113--281.
https://doi.org/10.1016/j.physrep.2008.04.005.

\bibitem{LiPLB2013}
Li, B.-A.; Han, X. Constraining the neutron-proton effective mass splitting using empirical constraints on the density dependence of nuclear symmetry
energy around normal density. {\it Phys. Lett. B} {\bf 2013},  \emph{727}, 276--281. https://doi.org/10.1016/j.physletb.2013.10.006.

\bibitem{Horowitz2014}
Horowitz, C.J.; Brown, E.F.; Kim, Y.; Lynch, W.G.; Michaels, R.; Ono, A.; Piekarewicz, J.; Tsang, M.B.; Wolter, H.H. A way forward in the study of the symmetry energy: experiment, theory, and observation. {\it J. Phys. G Nucl. Part. Phys.} {\bf 2014}, \emph{41}, 093001.
https://doi.org/10.1088/0954-3899/41/9/093001.

\bibitem{LattimerEPJA2014}
Lattimer, J.M.; Steiner, A.W. Constraints on the symmetry energy using the mass-radius relation of neutron stars. {\it Eur. Phys. J. A} {\bf 2014}, \emph{50}, 40.
https://doi.org/10.1140/epja/i2014-14040-y.

\bibitem{Drago2014}
Drago, A.; Lavagno, A.; Pagliara, G.; Pigato, D. Early appearance of $\Delta$ isobars in neutron stars. {\it Phys. Rev. C} {\bf 2014}, \emph{90}, 065809.
https://doi.org/10.1103/PhysRevC.90.065809.

\bibitem{Cai2015}
Cai, B.J.; Fattoyev, F.J.; Li, B.A.; Newton,W.G. Critical density and impact of $\Delta$(1232) resonance formation in neutron stars. {\it Phys. Rev. C} {\bf 2015}, \emph{92}, 015802.
https://doi.org/10.1103/PhysRevC.92.015802.

\bibitem{Zhu2016}
Zhu, Z.Y.; Li, A.; Hu, J.N.; Sagawa, H. $\Delta$(1232) effects in density-dependent relativistic hartree-fock theory and neutron stars. {\it Phys. Rev. C} {\bf 2016}, \emph{94}, 045803.
https://doi.org/10.1103/PhysRevC.94.045803.

\bibitem{Sahoo2018}
Sahoo, H.S.; Mitra, G.; Mishra, R.; Panda, P.K.; Li, B.A. Neutron star matter with $\Delta$ isobars in a relativistic quark model. {\it Phys. Rev. C} {\bf 2018}, \emph{98}, 045801.
https://doi.org/10.1103/PhysRevC.98.045801.

\bibitem{LiJJ2018}
Li, J.J.; Sedrakian, A.;Weber, F. Competition between delta isobars and hyperons and properties of compact stars. {\it Phys. Lett. B} {\bf 2018}, \emph{783}, 234--240.
https://doi.org/10.1016/j.physletb.2018.06.051.

\bibitem{LiJJ2019}
Li, J.J.; Sedrakian, A. Implications from GW170817 for Delta-isobar Admixed Hypernuclear Compact Stars. {\it Astrophys. J. Lett.} {\bf 2019}, \emph{874}, L22.
https://doi.org/10.3847/2041-8213/ab1090.

\bibitem{Ribes2019}
Ribes, P.; Ramos, A.; Tolos, L.; Gonzalez-Boquera, C.; Centelles, M. Interplay between $\Delta$ Particles and Hyperons in Neutron Stars.
{\it Astrophys. J.} {\bf 2019}, \emph{883}, 168. https://doi.org/10.3847/1538-4357/ab3a93.

\bibitem{Raduta2020}
Raduta, A.R.; Oertel, M.; Sedrakian, A. Proto-neutron stars with heavy baryons and universal relations. {\it Mon. Not. R. Astron. Soc.} {\bf2020}, \emph{499}, 914--931.
https://doi.org/10.1093/mnras/staa2491.

\bibitem{Raduta2021}
Raduta, A.R. $\Delta$-admixed neutron stars: Spinodal instabilities and dUrca processes. {\it Phys. Lett. B} {\bf 2021}, \emph{814}, 136070.
https://doi.org/10.1016/j.physletb.2021.136070.

\bibitem{Thapa2021}
Thapa, V.B.; Sinha, M.; Li, J.J.; Sedrakian, A. Massive $\Delta$-resonance admixed hypernuclear stars with antikaon condensations. 
{\it Phys. Rev. D} {\bf 2021}, \emph{103}, 063004. 
https://doi.org/10.1103/PhysRevD.103.063004.

\bibitem{Sen2021}
Sen, D. Variation of the $\Delta$ baryon mass and hybrid star properties in static and rotating conditions. {\it Phys. Rev. C} {\bf 2021}, \emph{103}, 045804.
https://doi.org/10.1103/PhysRevC.103.045804.

\bibitem{Jiang2012}
Jiang, W.Z.; Li, B.-A.; Chen, L.W. Large-mass neutron stars with hyperonization. {\it Astrophys. J.} {\bf 2012}, \emph{756}, 56. 

\bibitem{Providencia2019}
Provid\^{e}ncia, C.; Fortin, M.; Pais, H.; Rabhi, A. Hyperonic stars and the nuclear symmetry energy. {\it Front. Astron. Space Sci.} {\bf 2019}, \emph{6}, 13.
https://doi.org/10.3389/fspas.2019.00013.

\bibitem{Vidana2018}
Vida\~{n}a, I. Hyperons: The strange ingredients of the nuclear equation of state. {\it Proc. R. Soc. Lond. A} {\bf 2018}, \emph{474}, 20180145.
https://doi.org/10.1098/rspa.2018.0145.

\bibitem{Choi2021}
Choi, S.; Miyatsu, T.; Cheoun, M.K.; Saito, K. Constraints on Nuclear Saturation Properties from Terrestrial Experiments and
Astrophysical Observations of Neutron Stars. {\it Astrophys. J.} {\bf 2021}, \emph{909}, 156.
https://doi.org/10.3847/1538-4357/abe3fe.

\bibitem{Fortin2021}
Fortin, M.; Raduta, A.R.; Avancini, S.; Provid\^{e}ncia, C. Thermal evolution of relativistic hyperonic compact stars with calibrated
equations of state. {\it Phys. Rev. D} {\bf 2021}, \emph{103}, 083004.
https://doi.org/10.1103/PhysRevD.103.083004.

\bibitem{OzelPRD2010}
\"{O}zel, F.; Baym, G.; G\"{u}ver, T. Astrophysical Measurement of the Equation of State of Neutron Star Matter. {\it Phys. Rev. D} {\bf 2010}, \emph{82}, 101301.
https://doi.org/10.1103/PhysRevD.82.101301.

\bibitem{Steiner2010}
Steiner, A.W.; Lattimer, J.M.; Brown, E.F. The Equation of State from Observed Masses and Radii of Neutron Stars. {\it Astrophys. J.} {\bf 2010}, \emph{722}, 33.
https://doi.org/10.1088/0004-637X/722/1/33.

\bibitem{Steiner2013}
Steiner, A.W.; Lattimer, J.M.; Brown, E.F. The Neutron Star Mass-Radius Relation and the Equation of State of Dense Matter. {\it Astrophys. J. Lett.} {\bf 2013}, \emph{765}, L5.
https://doi.org/10.1088/2041-8205/765/1/L5.

\bibitem{Raithel2016}
Raithel, C.A.; \"{O}zel, F.; Psaltis, D. From Neutron Star Observables to the Equation of State. I. An Optimal Parametrization. {\it Astrophys. J.} {\bf 2016}, \emph{831}, 44.
https://doi.org/10.3847/0004-637X/831/1/44.

\bibitem{Raithel2017}
Raithel, C.A.; \"{O}zel, F.; Psaltis, D. From Neutron Star Observables to the Equation of State. II. Bayesian Inference of Equation of State Pressures.
{\it Astrophys. J.} {\bf 2017}, \emph{844},~156. https://doi.org/10.3847/1538-4357/aa7a5a.

\bibitem{Essick2020}
Essick, R.; Tews, I.; Landry, P.; Reddy, S.; Holz, D.E. Direct Astrophysical Tests of Chiral Effective Field Theory at Supranuclear Densities. {\it Phys. Rev. C} {\bf 2020}, 
\emph{102}, 055803. https://doi.org/10.1103/PhysRevC.102.055803.

\bibitem{Demorest2010}
Demorest, P.B.; Pennucci, T.; Ransom, S.M.; Roberts, M.S.; Hessels, J.W. A two-solar-mass neutron star measured using Shapiro delay. {\it Nature} {\bf 2010}, \emph{467}, 1081--1083.  
https://doi.org/10.1038/nature09466

\bibitem{Antoniadis2013}
Antoniadis, J.; Freire, P.C.; Wex, N.; Tauris, T.M.; Lynch, R.S.; Van Kerkwijk, M.H.; Kramer, M.; Bassa, C.; Dhillon, V.S.; Driebe, T.; et~al. A Massive Pulsar in a Compact Relativistic Binary. {\it Science} {\bf 2013}, \emph{340}, 6131. 
https://doi.org/10.1126/science.1233232.

\bibitem{Cromartie2020}
Cromartie, H.T.; Fonseca, E.; Ransom, S.M.; Demorest, P.B.; Arzoumanian, Z.; Blumer, H.; Brook, P.R.; DeCesar, M.E.; Dolch, T.; Ellis, J.A.; et~al. Relativistic Shapiro delay measurements of an extremely massive millisecond pulsar. {\it Nat. Astron.} {\bf 2020}, \emph{4}, 72--76. https://doi.org/10.1038/s41550-019-0880-2.

\bibitem{OzelApJ2016}
\"{O}zel, F.; Psaltis, D.; G\"{u}ver, T.; Baym, G.; Heinke, C.; Guillot. S. The Dense Matter Equation of State from Neutron Star Radius and Mass Measurements. 
{\it Astrophys. J.} {\bf 2016}, \emph{820}, 28. 
https://doi.org/10.3847/0004-637X/820/1/28.

\bibitem{Bogdanov2016}
Bogdanov, S.; Heinke, C.O.; \"{O}zel, F.; G\"{u}ver, T. Neutron Star Mass-Radius Constraints of the Quiescent Low-mass X-ray Binaries X7 and X5 in 
the Globular Cluster 47 Tuc. {\it Astrophys. J.} {\bf 2016}, \emph{831}, 184.
https://doi.org/10.3847/0004-637X/831/2/184.

\bibitem{Riley2019}
Riley, T.E.; Watts, A.L.; Bogdanov, S.; Ray, P.S.; Ludlam, R.M.; Guillot, S.; Arzoumanian, Z.; Baker, C.L.; Bilous, A.V.; Chakrabarty, D.; et~al. A NICER View of PSR J0030~+~0451: Millisecond Pulsar Parameter Estimation. {\it Astrophys. J. Lett.} {\bf 2019}, \emph{887}, L21.
https://doi.org/10.3847/2041-8213/ab481c.

\bibitem{Miller2019}
Miller, M.C.; Lamb, F.K.; Dittmann, A.J.; Bogdanov, S.; Arzoumanian, Z.; Gendreau, K.C.; Guillot, S.; Harding, A.K.; Ho, W.C.; Lattimer, J.M.; et~al. PSR J0030~+~0451 Mass and Radius from NICER Data and Implications for the Properties of Neutron Star Matter.
{\it Astrophys. J. Lett.}  {\bf 2019}, \emph{887}, L24. 
https://doi.org/10.3847/2041-8213/ab50c5.

\bibitem{BNS2017}
Abbott, B.; Jawahar, S.; Lockerbie, N.; Tokmakov, K. (LIGO Scientific Collaboration and Virgo Collaboration). GW170817: Observation of Gravitational Waves from a Binary Neutron Star Inspiral. 
{\it Phys. Rev. Lett.} {\bf 2017}, \emph{119}, 161101.

\bibitem{BNS2019}
Abbott, B.; Jawahar, S.; Lockerbie, N.; Tokmakov, K. (LIGO Scientific Collaboration and Virgo Collaboration). GW190425: Observation of a Compact Binary Coalescence with Total Mass 3.4 $M_{\odot}$.
{\it Astrophys. J. Lett.} {\bf 2020}, \emph{892}, L3.
https://doi.org/10.3847/2041-8213/ab75f5.

\bibitem{NSBH2021}
Abbott, R.; Abbott, T.D.; Abraham, S.; Acernese, F.; Ackley, K.; Adams, A.; Adams, C.; Adhikari, R.X.; Adya, V.B.; Affeldt, C.; et~al. Observation of Gravitational Waves from Two Neutron Star--Black Hole Coalescences. {\it Astrophys. J. Lett.} {\bf 2021}, \emph{915}, L5.
https://doi.org/10.3847/2041-8213/ac082e.

\bibitem{YagiSci2013}
Yagi, K.; Yunes, N. I-Love-Q. {\it Science} {\bf 2013}, \emph{341}, 365. https://doi.org/10.1126/science.1236462.

\bibitem{YagiPRD2013}
Yagi, K.; Yunes, N. I-Love-Q Relations in Neutron Stars and their Applications to Astrophysics, Gravitational Waves and Fundamental Physics, 
{\it Phys. Rev. D} {\bf 2013}, \emph{88}, 023009. 
https://doi.org/10.1103/PhysRevD.88.023009.

\bibitem{LeCun2015}
LeCun, Y.; Bengio, Y.; Hinton, G. Deep learning. {\it Nature} {\bf 2015}, \emph{521}, 436--444. https://doi.org/10.1038/nature14539

\bibitem{Goodfellow2016}
Goodfellow, I.; Bengio, Y.; Courville, A. {\it Deep Learning}; MIT Press: Cambridge, MA, USA, 2016. Available online: 
\url{https://www.deeplearningbook.org} (accessed on September 28, 2021).

\bibitem{He2016}
He, K.; Zhang, X.; Ren, S.; Sun, J. Deep residual learning for image recognition. In Proceedings of the IEEE Conference on Computer Vision 
and Pattern Recognition (CVPR), Las Vegas, NV, USA, 27--30 June 2016; pp. 770--778. 

\bibitem{Young2018}
Young, T.; Hazarika, D.; Poria, S.; Cambria, E. Recent trends in deep learning based natural language processing. {\it IEEE Comput. Intell. Mag.} {\bf 2018}, \emph{13}, 55--75.
https://doi.org/10.1109/MCI.2018.2840738.

\bibitem{Baker2019}
Baker, N.; Alexander, F.; Bremer, T.; Hagberg, A.; Kevrekidis, Y.; Najm, H.; Parashar, M.; Patra, A.; Sethian, J.; Wild, S.; et~al. {\it Workshop Report on Basic Research Needs for Scientific Machine Learning: Core Technologies for Artificial Intelligence}; Washington, DC, USA, 2019. https://doi.org/10.2172/1478744.

\bibitem{Pang2018}
Pang, L.G.; Zhou, K.; Su, N.; Petersen, H.; St\"{o}cker, H.; Wang, X.N. An equation-of-state-meter of quantum chromodynamics transition from deep learning. {\it Nat. Commun.} {\bf 2018}, \emph{9}, 210. https://doi.org/10.1038/s41467-017-02726-3.

\bibitem{Mori2017}
Mori, Y.; Kashiwa, K.; Ohnishi, A. Toward solving the sign problem with path optimization method. {\it Phys. Rev. D} {\bf 2017}, \emph{96}, 111501.
https://doi.org/10.1103/PhysRevD.96.111501.

\bibitem{Porotti2019}
Porotti, R.; Tamascelli, D.; Restelli, M.; Prati, E. Coherent transport of quantum states by deep reinforcement learning. {\it Commun. Phys.} {\bf 2019}, \emph{2}, 61. 
https://doi.org/10.1038/s42005-019-0169-x.

\bibitem{Rem2019}
Rem, B.S.; Käming, N.; Tarnowski, M.; Asteria, L.; Fläschner, N.; Becker, C.; Sengstock, K.; Weitenberg, C. Identifying quantum phase transitions using artificial neural networks on experimental data. {\it Nat. Phys.} {\bf 2019},
\emph{15}, 917--920. https://doi.org/10.1038/s41567-019-0554-0.

\bibitem{Melko2019}
Melko, R.G.; Carleo, G.; Carrasquilla, J.; Cirac, J.I. Restricted Boltzmann machines in quantum physics. {\it Nat. Phys.} {\bf 2019}, \emph{15}, 887--892. 
https://doi.org/10.1038/s41567-019-0545-1.

\bibitem{Carleo2017}
Carleo, G.; Troyer, M. Solving the Quantum Many-Body Problem with Artificial Neural Networks. {\it Science} {\bf 2017},\emph{ 355}, 602. 
https://doi.org/10.1126/science.aag2302.

\bibitem{Shanahan:2018vcv}
Shanahan, P.E.; Trewartha, D.; Detmold, W. Machine learning action parameters in lattice quantum chromodynamics.
{\it Phys. Rev. D} {\bf 2018}, \emph{97}, 094506. https://doi.org/10.1103/PhysRevD.97.094506.

\bibitem{Liu2021}
Liu, Z.; Tegmark, M. AI Poincar\`{e}: Machine Learning Conservation Laws from Trajectories. {\it Phys. Rev. Lett.} {\bf 2021}, \emph{126}, 180604.
https://doi.org/10.1103/PhysRevLett.126.180604.

\bibitem{Gomez:2021hqd}
Gomez, S.; Berger E.; Hosseinzadeh, G.; Blanchard, P.K.; Nicholl, M.; Villar, V.A. The Luminous and Double-peaked Type Ic Supernova 2019stc: 
Evidence for Multiple Energy Sources. {\it Astrophys. J.} {\bf 2021}, \emph{913}, 143. https://doi.org/10.3847/1538-4357/abf5e3.

\bibitem{Villar2020}
Villar, V.A.; Hosseinzadeh, G.; Berger, E.; Ntampaka, M.; Jones, D.O.; Challis, P.; Chornock, R.; Drout, M.R.; Foley, R.J.; Kirshner, R.P.; et~al. SuperRAENN: A Semisupervised Supernova Photometric Classification Pipeline Trained on Pan-STARRS1 Medium-Deep Survey Supernovae.
 {\it Astrophys. J.}  {\bf 2020}, \emph{905}, 94. https://doi.org/10.3847/1538-4357/abc6fd. 

\bibitem{Schwartz2021} 
Schwartz, M.D. Modern Machine Learning and Particle Physics. {\it Harv. Data Sci. Rev.} {\bf 2021}, Vol. \emph{3}, Issue 2. https://doi.org/10.1162/99608f92.beeb1183.

\bibitem{Gabbard2018}
Gabbard, H.; Williams, M.; Hayes, F.; Messenger, C. Matching Matched Filtering with Deep Networks for Gravitational-Wave Astronomy.
{\it Phys. Rev. Lett.} {\bf 2018}, \emph{120}, 141103. https://doi.org/10.1103/PhysRevLett.120.141103.

\bibitem{GeorgePRD2018}
George, D.; Huerta, E.A. Deep neural networks to enable real-time multimessenger astrophysics. {\it Phys. Rev. D} {\bf 2018}, \emph{97}, 044039.
https://doi.org/10.1103/PhysRevD.97.044039.

\bibitem{GeorgePLB2018}
George, D.; Huerta, E.A. Deep Learning for real-time gravitational wave detection and parameter estimation: Results with Advanced LIGO data.
{\it Phys. Lett. B} {\bf 2018}, \emph{778}, 64. https://doi.org/10.1016/j.physletb.2017.12.053.

\bibitem{Gebhard2019}
Gebhard, T.D.; Kilbertus, N.; Harry, I.; Sch\"{o}lkopf, B. Convolutional neural networks: A magic bullet for gravitational-wave detection? 
{\it Phys. Rev. D} {\bf 2019}, \emph{100}, 063015. https://doi.org/10.1103/PhysRevD.100.063015.

\bibitem{Wang2020}
Wang, H.; Wu, S.; Cao, Z.; Liu, X.; Zhu, J.Y. Gravitational-wave signal recognition of LIGO data by deep learning. {\it Phys. Rev. D} {\bf 2020}, \emph{101}, 104003.
https://doi.org/10.1103/PhysRevD.101.104003.

\bibitem{Lin:2020aps}
Lin, Y.C.; Wu, J.H.P. Detection of gravitational waves using Bayesian neural networks. {\it Phys. Rev. D} {\bf 2021}, \emph{103}, 063034.
https://doi.org/10.1103/PhysRevD.103.063034.

\bibitem{Morales2021}
Morales, M.D.; Antelis, J.M.; Moreno, C.; Nesterov, A.I. Deep Learning for Gravitational-Wave Data Analysis: A Resampling White-Box Approach. 
{\it Sensors} {\bf 2021}, \emph{21}, 3174. https://doi.org/10.3390/s21093174.

\bibitem{Xia2021}
Xia, H.; Shao, L.; Zhao, J.; Cao, Z. Improved deep learning techniques in gravitational-wave data analysis. {\it Phys. Rev. D} {\bf 2021}, \emph{103}, 024040.
https://doi.org/10.1103/PhysRevD.103.024040.

\bibitem{Chua2020}
Chua, A.J.K.; Vallisneri, M. Learning Bayesian Posteriors with Neural Networks for Gravitational-Wave Inference. {\it Phys. Rev. Lett.} {\bf 2020}, \emph{124}, 041102.
https://doi.org/10.1103/PhysRevLett.124.041102. 

\bibitem{Green2021}
Green, S.R.; Gair, J. Complete parameter inference for GW150914 using deep learning. {\it Mach. Learn. Sci. Technol.} {\bf 2021}, \emph{2}, 03LT01.
https://doi.org/10.1088/2632-2153/abfaed.

\bibitem{Wei2020}
Wei, W.; Huerta, E.A. Gravitational wave denoising of binary black hole mergers with deep learning. {\it Phys. Lett. B} {\bf 2020}, \emph{800}, 135081.
https://doi.org/10.1016/j.physletb.2019.135081.

\bibitem{Krastev2020}
Krastev, P.G. Real-time detection of gravitational waves from binary neutron stars using artificial neural networks. {\it Phys. Lett. B} {\bf 2020}, \emph{803}, 135330.
https://doi.org/10.1016/j.physletb.2020.135330.

\bibitem{Krastev2021}
Krastev, P.G.; Gill, K.; Villar, V.A.; Berger, E. Detection and parameter estimation of gravitational waves from binary neutron-star mergers in real 
LIGO data using deep learning. {\it Phys. Lett. B} {\bf 2021}, \emph{815}, 136161. https://doi.org/10.1016/j.physletb.2021.136161.

\bibitem{Lecun1998}
Lecun, Y.; Bottou, L.; Bengio, Y.; Haffner, P. Gradient-based learning applied to document recognition. {\it Proc. IEEE} {\bf 1998}, \emph{86}, 2278--2324.
https://doi.org/10.1109/5.726791.

\bibitem{Ferreira:2019bny}
Ferreira, M.; Provid\^encia, C. Unveiling the nuclear matter EoS from neutron star properties: a supervised machine learning approach. 
{\it J. Cos. Astropart. Phys.} {\bf 2021}, \emph{7}, 11. https://doi.org/10.1088/1475-7516/2021/07/011.

\bibitem{Morawski:2020izm}
Morawski, F.; Bejger, M. Neural network reconstruction of the dense matter equation of state derived from the parameters of neutron stars.
{\it Astron. Astrophys.} \textbf{2020}, 642, A78. https://doi.org/10.1051/0004-6361/202038130

\bibitem{Traversi:2020dho}
Traversi, S.; Char, P. Structure of Quark Star: A Comparative Analysis of Bayesian Inference and Neural Network Based Modeling.
\textit{Astrophys. J.} \textbf{2020}, \emph{905}, 9. https://doi.org/10.3847/1538-4357/abbfb4.

\bibitem{Fujimoto2020}
Fujimoto, Y.; Fukushima, K.; Murase, K. Mapping neutron star data to the equation of state using the deep neural network. \textit{Phys. Rev. D} \textbf{2020}, 
\emph{101}, 054016. https://doi.org/10.1103/PhysRevD.101.054016.

\bibitem{Stone2007}
Stone, J.R.; Reinhard, P.G. The Skyrme Interaction in finite nuclei and nuclear matter. \textit{Prog. Part. Nucl. Phys.} \textbf{2007}, 58, 587--657.
https://doi.org/10.1016/j.ppnp.2006.07.001.

\bibitem{Vautherin1972}
Vautherin, D.; Brink, D.M. Hartree-Fock Calculations with Skyrme's Interaction. I. Spherical Nuclei. \textit{Phys. Rev. C} \textbf{1972}, \emph{5},
626--647. https://doi.org/10.1103/PhysRevC.5.626.

\bibitem{Quentin1978}
Quentin, P.; Flocard, H. Self-Consistent Calculations of Nuclear Properties with Phenomenological Effective Forces. \textit{Annu. Rev. Nucl. Part. Sci.} \textbf{1978}, 
\emph{28}, 523--594. https://doi.org/10.1146/annurev.ns.28.120178.002515.

\bibitem{Boguta1977}
Boguta, J.; Bodmer, A.R. Relativistic calculation of nuclear matter and the nuclear surface. \textit{Nucl. Phys. A} \textbf{1977}, \emph{292}, 413--428.
https://doi.org/10.1016/0375-9474(77)90626-1.

\bibitem{Machleidt1987}
Machleidt, R.; Holinde, K.; Elster, C. The Bonn Meson Exchange Model for the Nucleon Nucleon Interaction. \textit{Phys. Rep.} \textbf{1987}, \emph{149}, 1--89.
https://doi.org/10.1016/S0370-1573(87)80002-9.

\bibitem{Nagelis1978}
Nagels, M.M.; Rijken, T.A.; de Swart, J.J. A Low-Energy Nucleon-Nucleon Potential from Regge Pole Theory. \textit{Phys. Rev. D} \textbf{1978},
\emph{17}, 768. https://doi.org/10.1103/PhysRevD.17.768.

\bibitem{Weinberg1990}
Weinberg, S. Nuclear forces from chiral lagrangians. \textit{Phys. Lett. B} \textbf{1990}, \emph{251}, 288--292. 
https://doi.org/10.1016/0370-2693(90)90938-3.

\bibitem{Weinberg1991}
Weinberg, S. Effective chiral lagrangians for nucleon-pion interactions and nuclear forces. \textit{Nucl. Phys. B} \textbf{1991}, \emph{363}, 3--18.
https://doi.org/10.1016/0550-3213(91)90231-L.

\bibitem{Epelbaum2009}
Epelbaum, E.; Hammer, H.W.; Mei{\ss}ner, U.G. Modern theory of nuclear forces. \emph{Rev. Mod. Phys.} \textbf{2009}, \emph{81}, 1773--1825.
https://doi.org/10.1103/RevModPhys.81.1773.

\bibitem{Day1967}
Day, B.D. Elements of the Brueckner-Goldstone Theory of Nuclear Matter. \textit{Rev. Mod. Phys.} \textbf{1967}, \emph{39}, 719--744. 

\bibitem{Brockmann1990}
Brockmann, R.; Machleidt, R. Relativistic nuclear structure. I. Nuclear matter. \textit{Phys. Rev. C} \textbf{1990}, \emph{42}, 1965--1980.
https://doi.org/10.1103/PhysRevC.42.1965.

\bibitem{Muther2017}
M\"uther, H.; Sammarruca, F.; Ma, Z. Relativistic effects and three-nucleon forces in nuclear matter and nuclei. \textit{Int. J. Mod. Phys. E}
\textbf{2017}, \emph{26}, 1730001. https://doi.org/10.1142/S0218301317300016.

\bibitem{Akmal1998}
Akmal, A.; Pandharipande, V.R.; Ravenhall, D.G. Equation of state of nucleon matter and neutron star structure. \textit{Phys. Rev. C}
\textbf{1998}, \emph{58}, 1804--1828. https://doi.org/10.1103/PhysRevC.58.1804.

\bibitem{Wiringa2000}
Wiringa, R.B.; Pieper, S.C.; Carlson, J.; Pandharipande, V.R. Quantum Monte Carlo calculations of A~=~8 nuclei. \textit{Phys. Rev. C} \textbf{2000},
\emph{62}, 014001. https://doi.org/10.1103/PhysRevC.62.014001.

\bibitem{Gandolfi2009}
Gandolfi, S.; Illarionov, A.Y.; Schmidt, K.E.; Pederiva, F.; Fantoni, S. Quantum Monte Carlo calculation of the equation of state of
neutron matter. \textit{Phys. Rev. C} \textbf{2009}, \emph{79}, 054005. https://doi.org/10.1103/PhysRevC.79.054005.

\bibitem{Kadanoff1962}
Kadanoff, L.; Baym, G. \textit{Quantum Statistical Mechanics}; W.A. Benjamin Inc.: New York, NY, USA, 1962. 
https://doi.org/10.1201/9780429493218.

\bibitem{Bogner2010}
Bogner, S.K.; Furnstahl, R.J.; Schwenk, A. From low-momentum interactions to nuclear structure. \textit{Prog. Part. Nucl. Phys.} \textbf{2010},
\emph{65}, 94--147. https://doi.org/10.1016/j.ppnp.2010.03.001.

\bibitem{Vidana2009}
Vida\~na, I.; Provid\^encia, C.; Polls, A.; Rios, A. Density dependence of the nuclear symmetry energy: A microscopic perspective. \textit{Phys. Rev. C} \textbf{2009},
80, 045806. https://doi.org/10.1103/PhysRevC.80.045806.

\bibitem{Zhang2018}
Zhang, N.-B.; Li, B.-A.; Xu, J. Combined Constraints on the Equation of State of Dense Neutron-rich Matter from Terrestrial Nuclear Experiments and 
Observations of Neutron Stars. \textit{Astrophys. J.} \textbf{2018}, \emph{859}, 90. https://doi.org/10.3847/1538-4357/aac027.

\bibitem{Cai2021}
Cai, B.J.; Li, B.A. Auxiliary Function Approach for Determining Symmetry Energy at Supra-saturation Densities. \textit{Phys. Rev. C}
\textbf{2021}, \emph{103}, 054611. https://doi.org/10.1103/PhysRevC.103.054611.

\bibitem{Zhang2019}
Zhang, N.B.; Li, B.A. Extracting Nuclear Symmetry Energies at High Densities from Observations of Neutron Stars and
Gravitational Waves. \textit{Eur. Phys. J. A} \textbf{2019}, \emph{55}, 39. https://doi.org/10.1140/epja/i2019-12700-0.

\bibitem{Xie2019}
Xie, W.J.; Li, B.A. Bayesian Inference of High-density Nuclear Symmetry Energy from Radii of Canonical Neutron Stars. 
\textit{Astrophys. J.} \textbf{2019}, \emph{883}, 174. https://doi.org/10.3847/1538-4357/ab3f37.

\bibitem{Krastev2006}
Krastev, P.G.; Sammarruca, F. Neutron star properties and the equation of state of neutron-rich matter. \textit{Phys. Rev. C}
\textbf{2006}, \emph{74}, 025808. https://doi.org/10.1103/PhysRevC.74.025808. 

\bibitem{Pethick1995}
Pethick, C.J.; Ravenhall, D.G.; Lorenz, C.P. The inner boundary of a neutron-star crust. \textit{Nucl. Phys. A} \textbf{1995}, \emph{584}, 675.
https://doi.org/10.1016/0375-9474(94)00506-I.

\bibitem{Haensel1994}
Haensel, P. ; Pichon, B. Experimental nuclear masses and the ground state of cold dense matter. \textit{Astron. Astrophys.} \textbf{1994}, 
\emph{283}, 313--318. 

\bibitem{Li:2020ass}
Li, B.A.; Magno, M. Curvature-slope correlation of nuclear symmetry energy and its imprints on the crust-core transition, 
radius and tidal deformability of canonical neutron stars. \textit{Phys. Rev. C} \textbf{2020}, \emph{102}, 045807.
https://doi.org/10.1103/PhysRevC.102.045807.

\bibitem{Oppenheimer1939}
Oppenheimer, J.R.; Volkoff, G.M. On Massive Neutron Cores. \textit{Phys. Rev.} \textbf{1939}, \emph{55}, 374. https://doi.org/10.1103/PhysRev.55.374.

\bibitem{Hinderer:2009ca}
Hinderer, T.; Lackey, B.D.; Lang, R.N.; Read, J.S. Tidal deformability of neutron stars with realistic equations of state and their gravitational 
wave signatures in binary inspiral. \textit{Phys. Rev. D} \textbf{2010},\emph{ 81}, 123016. https://doi.org/10.1103/PhysRevD.81.123016.

\bibitem{Flanagan:2007ix}
Flanagan, E.E.; Hinderer, T. Constraining neutron-star tidal Love numbers with gravitational-wave detectors. \textit{Phys. Rev. D} \textbf{2008}, \emph{77}, 021502.


\bibitem{Damour:2009vw}
Damour, T.; Nagar, A. Relativistic tidal properties of neutron stars. \textit{Phys. Rev. D} \textbf{2009}, \emph{80}, 084035.

\bibitem{Hinderer:2007mb}
Hinderer, T. Tidal Love Numbers of Neutron Stars. \textit{Astrophys. J.} \textbf{2008}, \emph{677}, 1216. https://doi.org/10.1086/533487.

\bibitem{Postnikov:2010yn}
Postnikov, S.; Prakash, M.; Lattimer, J.M. Tidal Love numbers of neutron and self-bound quark stars. \textit{Phys. Rev. D} \textbf{2010}, \emph{82}, 024016.
https://doi.org/10.1103/PhysRevD.82.024016.

\bibitem{Emmert-Streib2020}
Emmert-Streib, F.; Yang, Z.; Feng, H.; Tripathi, S.; Dehmer, M. An Introductory Review of Deep Learning for Prediction Models With Big Data.
\textit{Front. Artif. Intell.} \textbf{2020}, \emph{3}, 4. https://doi.org/10.3389/frai.2020.00004.

\bibitem{Neilsen2015}
Neilsen, M.A. \textit{Neural Networks and Deep Learning}; Determination Press: 2015. Available online: 
\url{http://neuralnetworksanddeeplearning.com}.

\bibitem{LeCun2012}
LeCun, Y.A.; Bottou, L.; Orr G.B.; M\"{u}ller, K.R. Efficient BackProp. In \textit{Neural Networks: Tricks of the Trade}; Lecture Notes in Computer Science; 
Springer: Berlin/Heidelberg, Germany, 2012; Volume 7700. https://doi.org/10.1007/978-3-642-35289-8\_3.

\bibitem{PREX:2021umo}
{Adhikari, D.; Albataineh, H.; Androic, D.; Aniol, K.; Armstrong, D.S.; Averett, T.; Gayoso, C.A.; Barcus, S.; Bellini, V.; Beminiwattha, R.S.; et~al. Accurate Determination of the Neutron Skin Thickness of $^{208}$Pb through Parity-Violation in Electron Scattering.
\textit{Phys. Rev. Lett.} \textbf{2021}, \emph{126}, 172502. https://doi.org/10.1103/PhysRevLett.126.172502}.

\bibitem{TF}
Abadi, M.; Agarwal, A.; Barham, P.; Brevdo, E.; Chen, Z.; Citro, C.; Corrado, G.S.; Davis, A.; Dean, J.; Dean, J.; et~al. TensorFlow: Large-Scale Machine Learning on Heterogeneous Distributed Systems. 2015. Available online: \url{https://www.tensorflow.org} (accessed on September 28, 2021).

\bibitem{Adam}
Kingma, D.P.; Ba, J. Adam: A method for stochastic optimization. \emph{arXiv} \textbf{2014}, arXiv:1412.6980.

\bibitem{Adam2}
Reddi, S.J.; Kale, S.; Kumar, S. On the convergence of Adam and beyond. \emph{arXiv} \textbf{2019}, arXiv:1904.09237.

\bibitem{Perreault2017}
Perreault Levasseur, L.; Hezaveh, Y.D.; Wechsler, R.H. Uncertainties in Parameters Estimated with Neural Networks: Application to Strong Gravitational Lensing.
\textit{Astrophys. J. Lett.} \textbf{2017}, \emph{850}, L7. https://doi.org/10.3847/2041-8213/aa9704.

\bibitem{Kobyzev2021}
Kobyzev, I.; Prince, S.J.D.; Brubaker, M.A.  Normalizing Flows: An Introduction and Review of Current Methods. \textit{IEEE Trans. Pattern Anal. Mach. Intell.} \textbf{2021}, \emph{43}, 3964--3979. https://doi.org/10.1109/TPAMI.2020.2992934.

\bibitem{Dax:2021tsq}
Dax, M.; Green, S.R.; Gair, J.; Macke, J.H.; Buonanno, A.; Sch\"olkopf, B. Real-time gravitational-wave science with neural posterior estimation. \emph{arXiv} \textbf{2021}, arXiv:2106.12594

\bibitem{LIGOScientific:2018mvr}
Abbott, B.P.; Abbott, R.; Abbott, T.; Abraham, S.; Acernese, F.; Ackley, K.; Adams, C.; Adhikari, R.X.; Adya, V.B.; Affeldt, C.; et~al. GWTC-1: A Gravitational-Wave Transient Catalog of Compact Binary Mergers Observed by LIGO and Virgo during 
the First and Second Observing Runs. \textit{Phys. Rev. X} \textbf{2019}, \emph{9}, 031040. https://doi.org/10.1103/PhysRevX.9.031040.
\end{references}
\end{document}